\documentclass[aps,pre,preprint,superscriptaddress]{revtex4-1}
\usepackage{bm}
\usepackage{graphicx}
\usepackage{amsmath}
\usepackage[english]{betababel}

\usepackage{amssymb}
\usepackage{latexsym}

\usepackage{url}
\usepackage{xcolor}
\definecolor{newcolor}{rgb}{.8,.349,.1}

\usepackage{bm}
\usepackage{mathtools}
\usepackage{amsmath}
\usepackage{amssymb}
\usepackage{xcolor}
\usepackage{booktabs}
\usepackage{multirow}
\usepackage{adjustbox}
\usepackage{mathrsfs}

\begin{document}

\title{A numerical method for inextensible elastic filaments in viscous fluids}%

\author{Mehdi Jabbarzadeh}%
\email{jabbarzadeh.mehdi@utah.edu}
\affiliation{
	Department of Mechanical Engineering, University of Utah, Salt Lake City, UT 84112 USA}
\author{Henry Chien Fu}%
\affiliation{
	Department of Mechanical Engineering, University of Utah, Salt Lake City, UT 84112 USA}

\begin{abstract}
The deformations of flagella are important in the motility of single- and multi-flagellated bacteria. \textcolor{black}{Existing numerical methods have treated flagella as extensible filaments with a large extensional modulus, resulting in a stiff numerical problem and long simulation times.  However, flagella are nearly inextensible, so to avoid large extensional stiffness}, we introduce inextensible elastic rod models with hydrodynamics treated by a surface distribution of regularized Stokeslets.
We benchmark this new model against previously described models of extensible elastic rods with hydrodynamics treated by a centerline distribution of regularized Stokeslets and rotlets, as well as a surface distribution of regularized Stokeslets. 
We compare the accuracy of the inextensible model with the extensible models and illustrate for which ratios of stretching/bending stiffness (which depend on the diameter of filament) the inextensible model is accurate and more efficient. We show that our inextensible approach can be markedly more efficient than the extensible models for many biological filaments. 
We also compare the accuracy of the centerline distribution of the Stokeslets and rotlets to the more accurate surface distribution of Stokeslets for modeling fluid-structure interactions of filaments.
\end{abstract}

\maketitle


 
\section{Introduction} \label{sec:intro}
The fluid-structure interactions of elastic filaments are important for many biological applications. 
Important examples include the dynamics of 
elastic flagella or cilia which are involved in microorganism locomotion \cite{Vogel2010,Berg,Jabbarzadeh2018,Funfak2014,Ali2017,Ohmura2018,Nishigami2018,Constantino2018,Ahmadvand2019}, 
biofilm streamers which influence ecosystem processes \cite{Rusconi2011,Drescher2013,DuRoure2019}, 
flexible microtubule and motor protein assemblies which occur in cytoskeletal networks
\cite{Memet2018,Gittes1993,Kikumoto2006,Fujime1972,Gittes1993,Jia2017}, 
and super-coiled DNA \cite{Schlick1995,Schlick1992,Benham1979,Shi1994}. 
Typically at these small scales the dynamics, deformations, and interactions of filaments with each other and surrounding fluid occur in the Stokes limit where viscous forces are dominant and inertial effects are negligible.

Experimental investigations show a wide range of rigidity for these different flexible filaments. In Table \ref{table:parameters}, we have summarized the mechanical properties and typical diameter of the most common filaments in viscous flows. 
The Young's modulus $E$ that measures the stiffness of these filaments varies over a wide range of $10^4-10^{10} ~ \rm pN /\mu m^2$. The rigidity of a filament with diameter $d$ can be characterized by the bending rigidity ${EI}\sim E d^4$ and stretching rigidity ${EA}\sim Ed^2$, which depend on the cross sectional second moment of area $I=\pi d^4/64$ or area $A=\pi d^2/4$. For filaments with homogeneous material properties and circular cross sections, the ratio between stretching and bending stiffness only depends on diameter of the filaments as ${EA/EI} \sim d^{-2}$. For smaller diameters this ratio becomes  large meaning that stretching is much less important than bending of thin filaments. In Table \ref{table:parameters}, we also report the ratio of ${EA/EI}$ for different filaments in viscous flow which can be in the range of $10^2-10^5\rm \mu m^{-2}$. 

Different mathematical and computational methods have been developed to describe the interaction of flexible filaments with Stokesian fluids. 
\textcolor{black}{
	Many previous approaches use Euler-Bernoulli beam theory to describe intrinsically straight filaments interacting with external force densities from the surrounding fluid \cite{Spagnuolo2019,Gazzola2016,DuRoure2019,Tornberg2004,Chakrabarti2019,Liu2018,Manikantan2015,Chakrabarti2019a}.
	However, since Euler-Bernoulli theory ignores torsional elasticity for 3D deformations, here we focus on methods that 
	employ the Kirchhoff rod model as described in \S \ref{sec:Kirchhoff} to describe the mechanics of a slender filament. The Kirchhoff rod theory also readily accounts for filaments with intrinsic curvature such as bacterial flagella. In these methods
%
%
the mechanics of the filament} have been coupled to fluid flow using various approaches such as slender body theory \cite{Lighthill1976,Keller1976,Nazockdast2017,Jabbarzadeh2014}, boundary integral methods \cite{Phan-Thien1987,Greengard2004,Shum2010,Jabbarzadeh2018a}, immersed boundary methods \cite{Bringley2008,Stein2019}, and the method of regularized Stokeslets \cite{Cortez2001,Cortez2005,Olson2013,Martindale2016,Constantino2016,Fu2015,Samsami2020} using both centerline and surface distributions of regularized Stokeslets.
While slender body theory and centerline distributions of regularized Stokeslets are fast and easy to implement, they can be inaccurate for near field interactions \cite{Martindale2016}. On the other hand, the boundary integral and immersed boundary methods can more accurately simulate fluid flow, but they are computationally expensive and more complicated to implement \cite{Shum2010,Jabbarzadeh2018a}. 
In this paper, we use surface distributions of regularized Stokeslets to characterize the performance of our inextensible rod model since they remain easy to implement while allowing accurate resolution of fluid flows.  

\begin{table}[!t]
	\centering
	\caption{Mechanical properties of flexible filaments in viscous flow.}
	\begin{adjustbox}{width=\textwidth,center=\textwidth}
		\begin{tabular}{@{}ccccccccc@{}}
			\toprule
			Filament &  $EI$           & $GJ$           & $E$            & $G$    & $d$ & $\textcolor{black}{L}$ & $EA/EI$  &  References \\ 
			type     & $(\rm pN \mu m^2)$ & $(\rm pN \mu m^2)$ & $(\rm pN/\mu m^2)$ & $(\rm pN\mu m^2)$ & $(\rm nm)$ &$\textcolor{black}{(\mu m)}$ & $(\rm 1/\mu m^2)$  \\ \midrule
			Prokaryotic Flagella & $1-3.5$ & $4.6$ & $\sim \! \! 10^{8}-10^{10}$ & $\sim \! \! 10^9$ & $10-25$ & \textcolor{black}{$3-25$} & $\sim \! \! 10^4$ & \cite{Darnton2007,Fujime1972,Hoshikawa1985,Takano2005}           \\
			Eukaryotic flagella/Cilium &$300-4000$ &  $ $ &   & $ $  & $200$ & \textcolor{black}{$5-50$} & $\sim \! \! 10^2$ &   \cite{Xu2016,Satir2007a,Nicastro2005} \\
			Hook & $0.0002-0.2$ &  $0.002-0.4$ & $\sim \! \! 10^4-10^6$ & & 9 & \textcolor{black}{$0.1$} & $\sim \! \! 10^4$ &    \cite{Son2013}  \\
			Microtubules & $2-8$ &  & $\sim \! \! 10^9$  &   & $10-20$ & \textcolor{black}{$3-100$} & $\sim \! \! 10^4$ & \cite{Memet2018,Gittes1993,Kikumoto2006}  \\ 
			Actin filaments & $0.03-0.16$ &   & $\sim \! \! 10^9$  & & $6-10$ & \textcolor{black}{$4-30$} & $\sim \! \! 10^5$ & \cite{Fujime1972,Gittes1993,Jia2017}\\    \bottomrule
		\end{tabular}
	\end{adjustbox}
	\label{table:parameters}
\end{table}

\subsection{Characteristic length and timescales of filaments in viscous flow} \label{sec:timescale}

\begin{figure}[t]
	\centering
	\includegraphics[width=0.4\linewidth]{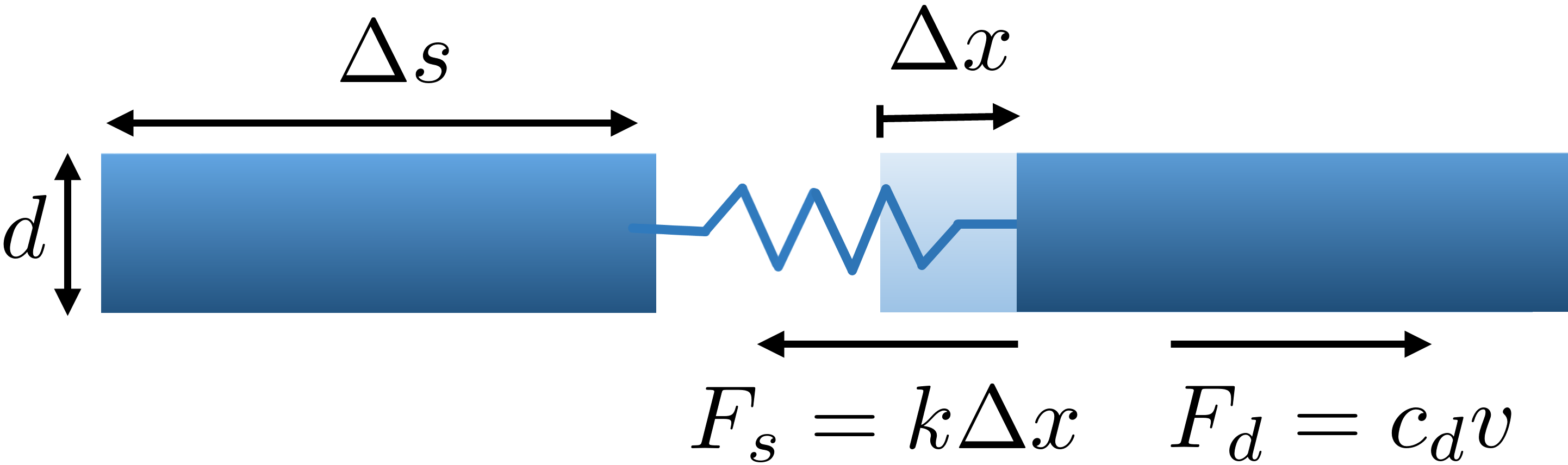}
	\caption{Simplified example to estimate timescales $\tau$ needed to resolve dynamics for numerical approaches. Two rigid cylindrical segments of length $\Delta s$ are connected by a linear spring $k$ and immersed in a viscous flow. A small relative displacement ($\Delta x$) between two segments leads to spring and hydrodynamic forces of $F_s=k \Delta x$ and $F_d=\textcolor{black}{c_d} v$, respectively, where \textcolor{black}{$c_d$} is the drag coefficient of a cylindrical segment and $v$ is the velocity. 
	Force balance determines the velocity $v=k/\textcolor{black}{c_d} \Delta x$, and resolving the dynamics of the problem requires timesteps $ v \delta t \ll \Delta x$, or $\delta t \ll \tau = \textcolor{black}{c_d}/k$.
	}
	\label{fig:timescaleexample}
\end{figure}

The computational expense of numerical approaches to stiff problems is determined by the size of the time-step needed  
to resolve the dynamics.
For fluid-structure interactions of slender filaments, the maximum time-step (hence least expensive computation) is given by characteristic timescales that can be estimated as follows.
In numerical approaches an elastic slender filament of diameter $d$ is usually discretized into small cylindrical segments of length $\Delta s$. The constitutive laws of the Kirchhoff model determine the internal forces and torques transmitted through cross-sections due to shape deformations. For this purpose consider the discretized segments as attached to each other by bending and stretching springs with strengths that can be estimated from the Kirchhoff rod model. For simplicity, consider just two rigid cylindrical segments of length $\Delta s$ and diameter $d$ connected by a linear spring (spring stiffness of $k$) as shown in Fig. \ref{fig:timescaleexample}. For a small relative displacement of $\Delta x$ between these two segments, the forces applied on segments are the spring ($F_s=k \Delta x$) and drag forces ($F_d=c_d v$) where \textcolor{black}{$c_d$} is the drag coefficient and $v$ is the velocity of segments. The Stokes limit implies force balance on the segments, so one can calculate the velocity as $v=(k/\textcolor{black}{c_d}) \Delta x$. To resolve the dynamics, we need a time-step $\delta t$ such that $v \delta t  \ll \Delta x$, or $\delta t \ll \textcolor{black}{c_d}/k = \tau$, where $\tau$ is the characteristic timescale of the problem. 

\begin{figure}[b]
	\centering
	\includegraphics[width=0.6\linewidth]{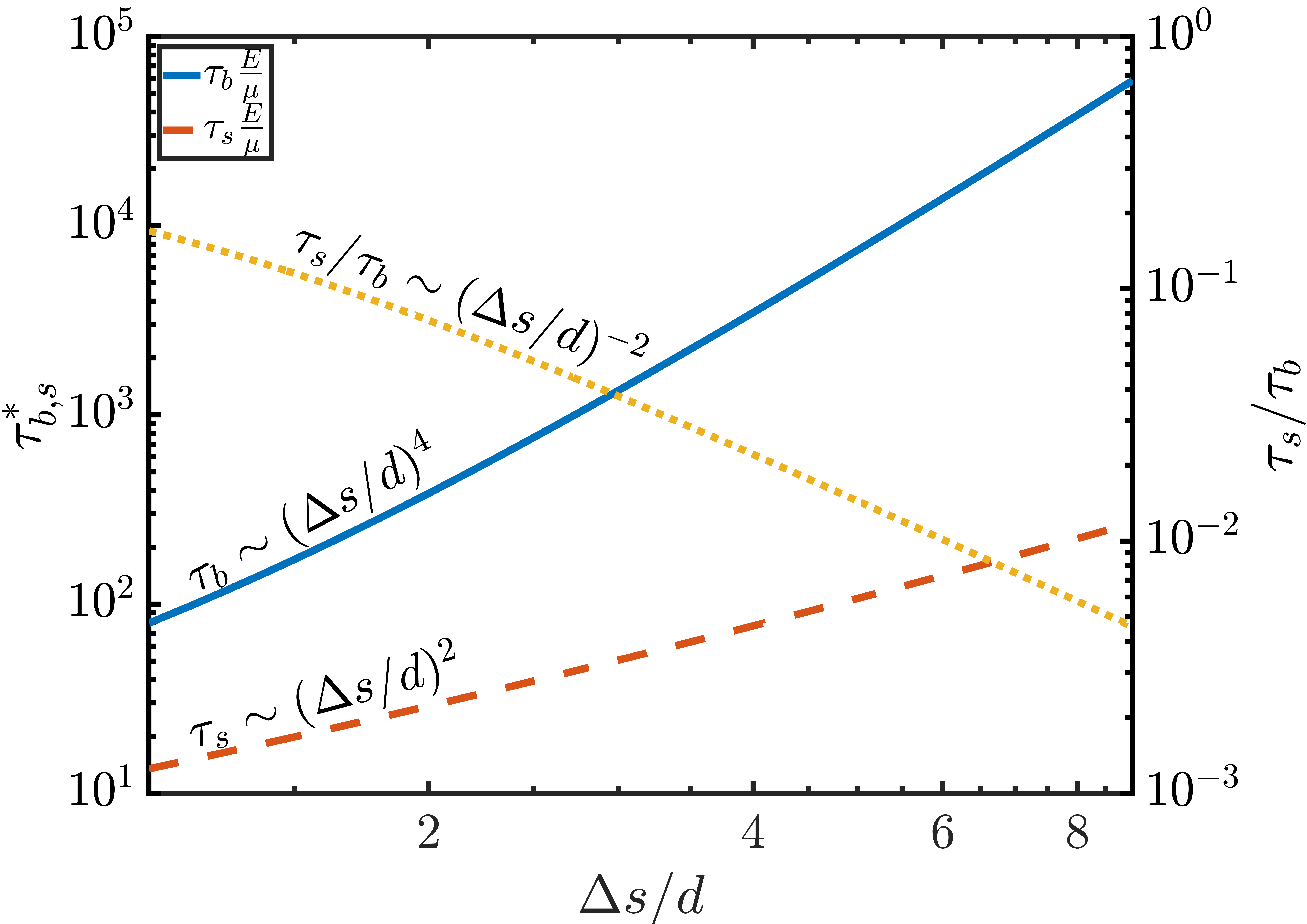}
	\caption{Nondimensional stretching and bending timescales as a function of segment size $\Delta s$ normalized by the diameter of filament $d$. $E$ is the Young's modulus and $\mu$ is the viscosity of surrounding fluid. The stretching timescale is smaller than the bending timescale and limits the speed of simulations.}
	\label{fig:timescale}
\end{figure}

To estimate relevant timescales for the fluid-structure interactions of filaments, we estimate that hydrodynamic resistances for cylindrical segments in viscous flow  for the translational and rotational motions scale as $\textcolor{black}{c_s} \sim \mu \Delta s$ and $\textcolor{black}{c_b} \sim \mu {\Delta s} ^3$, respectively, where $\mu$ is the  viscosity of the surrounding fluid.  
The stretching spring constant and bending spring constant are $k_s \sim {EA}/\Delta s$ and $k_b \sim {EI}/\Delta s$, respectively, where $E$ is the Young's modulus, and $A$ and $I$ are the cross-sectional area and second moment of area. Thus, the stretching and bending timescales can be estimated as $\tau_s=\mu(\Delta s/d)^2 /E$ and $\tau_b=\mu (\Delta s/d)^4/E$, respectively. 
Since smaller timesteps lead to longer computational times, in numerical simulations the overall solution time is controlled by the smallest stretching or bending timescale.  
In Fig. \ref{fig:timescale}, we plot non-dimensional timescales for the stretching and bending of filaments  ($\tau_{s,b}^*=(E/\mu) \tau_{s,b}$) as a function of segment aspect ratios $(\Delta s/d)$. 
The ratio of the stretching timescale to the bending timescale is $\tau_s/\tau_b = {EI}/({EA}~{\Delta s}^2)\sim (\Delta s/d)^{-2}$. In Fig. \ref{fig:timescale} the stretching timescale is at least one order of magnitude smaller than the bending timescale and this difference increases as $\Delta s/d$ increases, i.e, as the filament becomes thinner. This analysis shows that the stretching timescale, which is related to the extensibility of filaments, determines the solution times in the numerical approaches. 

In this paper, we will introduce a new numerical treatment of fluid-structure interaction of filaments in viscous flow using the inextensible Kirchhoff rod model.  Imposing inextensible constraints eliminates stretching timescales, so that only bending timescales remain, allowing the use of larger time steps to shorten simulation times. 
We characterize when our inextensible model is more efficient than extensible models. \textcolor{black}{To determine when the inextensible approximation is valid for any biological filaments, which are physically extensible (even if only slightly), we also compare the new inextensible model with existing extensible solutions.}

\section{Methods}
For all the methods used in this paper, the general approach is to take the current configuration of the filament, discretize its centerline into straight segments, and use a Kirchhoff rod theory to 
calculate the internal stresses from a given configuration by comparing the deformed shape to an equilibrium configuration as explained in \S \ref{sec:Kirchhoff}.   From the internal stresses and force/torque balance, one can then calculate the external hydrodynamic forces on segments, which are equal and opposite to the forces exerted on the fluid by the filament.  In this paper, we use the method of regularized Stokeslets to find the fluid velocity field from the forces exerted on the fluid, and then the no-slip condition determines the velocities of segments from the fluid flow.

\begin{figure}[]
	\centering
	\includegraphics[width=0.6\linewidth]{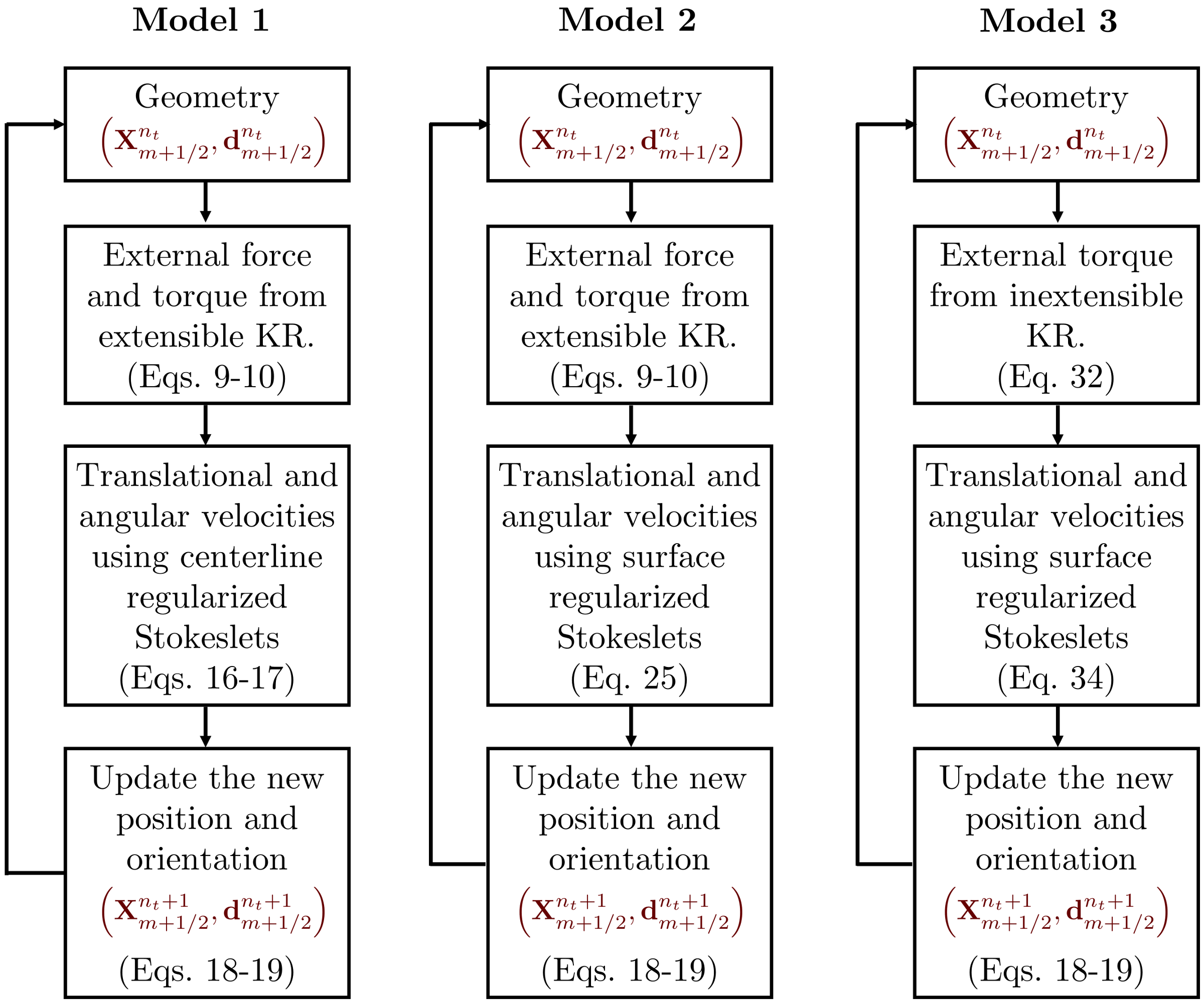}
	\caption{Schematics of algorithms for different numerical models used in this paper. For model 1, we use an extensible version of the Kirchhoff (KR) rod theory to calculate the external forces and torques and a centerline distribution of regularized Stokeslets to describe the hydrodynamics. For model 2, we use the extensible version of Kirchhoff rod theory and a surface distribution of regularized Stokeslets. For model 3, we use an inextensible Kirchhoff rod theory and a surface distribution of regularized Stokeslets to describe hydrodynamic interactions.}
	\label{fig:Algorithm}
\end{figure} 

Here, we explain and compare three different approaches and numerical procedures used in this paper (Fig. \ref{fig:Algorithm}). 
In the first model, which was previously developed and used in \cite{Olson2013,Park2017,Carichino2018,Park2019}, the hydrodynamic interactions are described by a centerline distribution of regularized Stokeslets, in which each segment is represented by a regularized Stokeslet and rotlet pair to describe the hydrodynamics of that segment. The internal forces and torques in the structure are calculated by an extensible version of the Kirchhoff rod model. 
In the second model, which we previously used in \cite{Jabbarzadeh2018},  we use a distribution of regularized Stokeslets on the surface of segments to more accurately satisfy boundary conditions in the calculation of hydrodynamic interactions, but like the first model we use the extensible Kirchhoff rod to describe the filament mechanics.
In the rest of the manuscript, we refer to the second model as the ``extensible model,'' and the first model by explicitly specifying the use of a centerline distribution.
In the third model, which is the main contribution of this paper, hydrodynamic interactions are also treated by a surface distributions of regularized Stokeslets, but we impose inextensibility condition on the Kirchhoff rod model to avoid stretching timescales and decrease the number of degrees of freedom which speeds up simulation times. 

\subsection{Kirchhoff  rod model} \label{sec:Kirchhoff}
\begin{figure}[t]
	\centering
	\includegraphics[width=0.6 \linewidth]{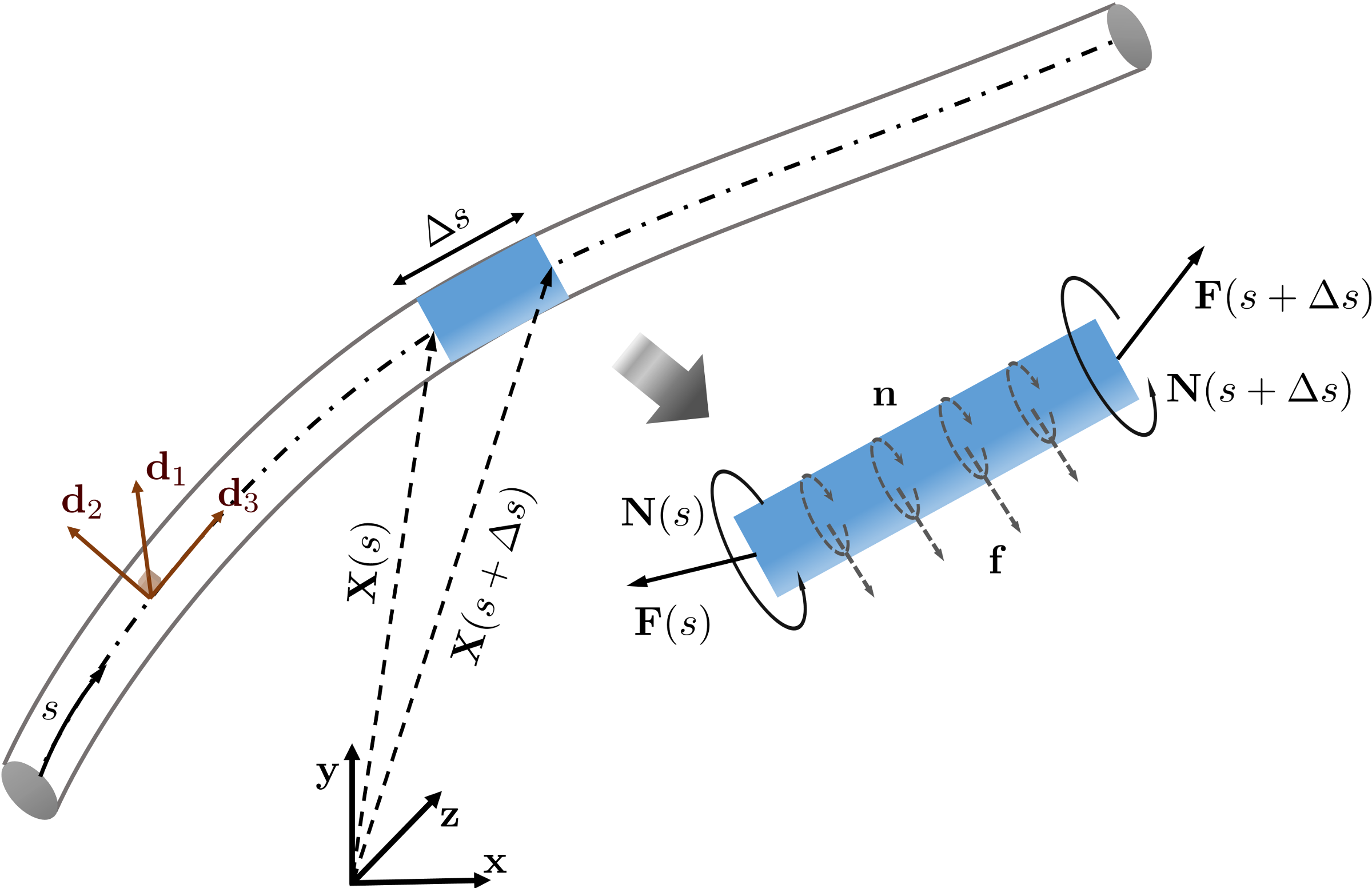}
	\caption{Kirchhoff rod theory is used to describe the mechanics of the slender filaments. The position of the centerline of the rod (dash-dots) is specified by $\mathbf{X}(s)$ where $s$ is a Lagrangian parameter along the arclength. Orthonormal triads $\{\mathbf{d}^1 (s),\mathbf{d}^2 (s),\mathbf{d}^3 (s)\}$ define the orientation of material cross-sections on the centerline.  $\mathbf{F}(s)$ and $\mathbf{N}(s)$ are the force and torque exerted across cross sections by the material with greater s on material with lesser s, while $\mathbf{f}(s)$ and $\mathbf{n}(s)$ are external force and torque (linear) densities exerted by the fluid on the filament.}
	\label{fig:rod}
\end{figure}

Since the filaments are slender, we adopt a version of the Kirchhoff rod model described by \cite{Olson2013, Lim2008} to deal with the flexibility of extensible and inextensible filaments at low Reynolds number. In the standard Kirchhoff rod model, for a homogeneous and isotropic rod, the centerline of a filament is described by a space curve $\mathbf{X}(s)$ where $s\in[0,L]$ is a Lagrangian parameter along the arclength and $L$ is the length of filament (Fig. \ref{fig:rod}). 
There is a set of orthonormal basis vectors  $\{\mathbf{d}^1 (s),\mathbf{d}^2 (s),\mathbf{d}^3 (s)\}$ associated with the material of each cross-section, where $\mathbf{d}^3$ is normal to the cross-section in the direction of positive $s$.
The internal force and torque exerted by material with larger $s$ on material with smaller $s$ through the cross-section at $s_0$ are represented by $\mathbf{F}(s_0)$ and $\mathbf{N}(s_0)$, respectively. In the presence of external force $\mathbf{f}(s)$ and torque $\mathbf{n}(s)$ density exerted by the fluid on the filament, the Kirchhoff equations for the force and torque are given by  
\begin{eqnarray}
\frac{\partial \mathbf{F}}{\partial s} + \mathbf{f}& =& \mathbf{0}, \label{forcebalance} \\   
\frac{\partial \mathbf{N}}{\partial s} + \mathbf{d}^3\times\mathbf{F} + \mathbf{n}&=& \mathbf{0}. \label{torquebalance}
\end{eqnarray} 
The forces and torques in Eqs. (\ref{forcebalance}) and (\ref{torquebalance}) are expanded in the $\mathbf{d}^i$ basis as 
\begin{eqnarray}
\mathbf{F} &=& \sum_{i=1}^{3}{F^i}\mathbf{d}^i, \qquad \mathbf{N} = \sum_{i=1}^{3}{N^i}\mathbf{d}^i, \\
\mathbf{f} &=& \sum_{i=1}^{3}{f^i}\mathbf{d}^i, \qquad \mathbf{n} = \sum_{i=1}^{3}{n^i}\mathbf{d}^i.
\end{eqnarray} \label{eq5}
The constitutive relations specifying the torques in the extensible Kirchhoff rod model are given by
\begin{eqnarray}
N^1 &=& {EI} (\frac{\partial \mathbf{d}^2} {\partial s}\cdot \mathbf{d}^3-\bm{\nu}^1 ), \nonumber \\ 
N^2 &=& {EI} (\frac{\partial \mathbf{d}^1} {\partial s}\cdot \mathbf{d}^3-\bm{\nu}^2 ), \nonumber \\ 
N^3 &=& {GJ} (\frac{\partial \mathbf{d}^2} {\partial s}\cdot \mathbf{d}^1-\bm{\nu}^3 ), \label{eq:rodtorque}
\end{eqnarray}
and specifying the forces are
\begin{eqnarray}
F^1 &=& {GA} (\mathbf{d}^1 \cdot \frac{\partial \mathbf{X}}{\partial s}), \nonumber \\   
F^2 &=& {GA} (\mathbf{d}^2 \cdot \frac{\partial \mathbf{X}}{\partial s}), \nonumber \\   
F^3 &=& {EA} (\mathbf{d}^3 \cdot \frac{\partial \mathbf{X}}{\partial s}-1).   \label{eq:rodforce}
\end{eqnarray}
\textcolor{black}{where $J$ is the second polar moment of area ($J=2I$ for circular cross-sections), and $G$ is the shear modulus. In this paper, we assume $G=E$ following previous works \cite{Shum2010,Olson2013,Jabbarzadeh2018,Park2017,Park2019}.} The intrinsic curvature $\kappa=\sqrt{\nu_1^2+\nu_2^2}$ and intrinsic twist vector $\nu_3$ are defined by the strain twist vector $\bm{\nu} = (\nu_1,\nu_2,\nu_3)$, and allow us to specify the undeformed shape of the filament.

We apply the Kirchhoff rod model to find the internal and external forces and torques for a known filament configuration.
We discretize the centerline of filament into $M$ segments along the arclength with equal lengths of $\Delta s$.  $M+1$ material points are labeled by an integer index $m=0,1,2,\cdots,M$, such that $s_m=m\Delta s$ represents the arclength coordinate of each point along the centerline. Segments between these points are assumed to be straight cylinders with diameter  ${d}$ and length $\Delta s$ moving as rigid bodies. The center of cylinders is specified by half-integer values $s_{m+1/2}$.   Positions and basis vectors at segment endpoints and centers are defined by $\mathbf{X}_{n} = \mathbf{X}(s_{n})$ and $\mathbf{d}^i_{n} = \mathbf{d}^i(s_n)$, for n integer or half-integer. For a given configuration, we calculate the internal torques transmitted between cross-sections at material points $s_m$ using the discrete form of the constitutive equations for the torque (Eq. (\ref{eq:rodtorque})),
\begin{eqnarray}
\mathbf{N}_m &=& \sum_{i=1}^{3}{N^i_m}\mathbf{d}^i_m\nonumber\\
N^1_m &=& {EI} (\frac{\mathbf{d}^2_{m+1/2} - \mathbf{d}^2_{m-1/2}} {\Delta s}\cdot \mathbf{d}^3_m-\bm{\nu}^1_m ),\nonumber \\ 
N^2_m &=& {EI} (\frac{\mathbf{d}^1_{m+1/2} - \mathbf{d}^1_{m-1/2}} {\Delta s}\cdot \mathbf{d}^3_m-\bm{\nu}^2_m ),\nonumber \\ 
N^3_m &=& {GJ} (\frac{\mathbf{d}^2_{m+1/2} - \mathbf{d}^2_{m-1/2}} {\Delta s}\cdot \mathbf{d}^1_m-\bm{\nu}^3_m ), \label{eq:rodforce_dis}
\end{eqnarray}
and calculate the internal forces from the discrete form of Eq. \ref{eq:rodforce},
\begin{eqnarray}
\mathbf{F}_m &=& \sum_{i=1}^{3}{F^i_m}\mathbf{d}^i_m\nonumber\\
F^1_m &=& {GA} (\frac{\mathbf{X}_{m+1/2}-\mathbf{X}_{m-1/2}}{\Delta s} \cdot \mathbf{d}^1_m),   \nonumber \\
F^2_m &=& {GA} (\frac{\mathbf{X}_{m+1/2}-\mathbf{X}_{m-1/2}}{\Delta s} \cdot \mathbf{d}^2_m ),   \nonumber \\
F^3_m &=& {EA} (\frac{\mathbf{X}_{m+1/2}-\mathbf{X}_{m-1/2}}{\Delta s} \cdot \mathbf{d}^3_m-1). \label{eq:rodtorque_dis}
\end{eqnarray}
Then, we compute the external force and torque applied on the center of each segment ($s_{m+1/2}$) using the discrete form of Eqs. (\ref{forcebalance}) and (\ref{torquebalance}),
\begin{eqnarray}
\mathbf{N}_{m+1/2} &=& \mathbf{N}_{m} - \mathbf{N}_{m+1} - \frac{\Delta s}{2}(\mathbf{d}^3_{m+1}\times \mathbf{F}_{m+1} + \mathbf{d}^3_{m}\times \mathbf{F}_{m}),\label{eq:torquebalance_dis} \\   
\mathbf{F}_{m+1/2} &=& \mathbf{F}_{m} - \mathbf{F}_{m+1}.   \label{eq:forcebalance_dis}
\end{eqnarray}
Here $\mathbf{F}_{m+1/2}= \mathbf{f}_{m+1/2} \Delta s$ and $\mathbf{N}_{m+1/2} = \mathbf{n}_{m+1/2} \Delta s $ are equal to the total forces and torques applied from surrounding fluid on the segment $s_m$. 

\subsection{Hydrodynamic interactions determine velocities and rotation rates along the filament} \label{sec:numerical}
In \S \ref{sec:Kirchhoff}, we described how the deformed filament geometry determines the external forces and moments applied by the fluid on segments of a discretized filament.  In this section, we explain how these forces and moments can be used to calculate the hydrodynamic flow fields which determine the velocities and rotation rates of segments.   To deal with the hydrodynamics, we use the method of regularized Stokeslets \cite {Cortez2001,Cortez2005,Martindale2016}.
In \S \ref{sec:cnt} we describe the use of a centerline distribution of regularized Stokeslets, while in \S \ref{sec:ext} we describe the use of a surface discretization of regularized Stokeslets.

\subsubsection{Method 1: Extensible Kirchhoff rod and centerline distribution of regularized Stokeslets } \label{sec:cnt}
Given the filament shape, the forces and moments exerted by the fluid on the segments are calculated as in \S \ref{sec:Kirchhoff}. 
In the zero Reynolds numbers limit, viscous forces dominate and the fluid is governed by the Stokes equation. In the method of regularized Stokeslets, the fundamental solution is obtained for a volumetric force $\mathbf{g}^b$ exerted by the fluid onto a spherical blob with radius $\epsilon$ instead of at a singular point \cite{Cortez2001, Cortez2005, Cortez2005a,Martindale2016}. The Stokes equation and incompressibility of the fluid can be expressed as
\begin{eqnarray}
-\nabla p + \mu \nabla^2 \mathbf{u} - \mathbf{g}^b &=& 0, \label{eq:stokes} \\   
\nabla \cdot \mathbf{u} &=& 0. \label{eq:cont}
\end{eqnarray} 
where $\mu$ is the fluid viscosity, $\mathbf{u}$ is the fluid velocity, and $\mathbf{g}^b$ is given by $\mathbf{g}^b(\mathbf{x}) = \bm{\mathcal{F}} \psi_\epsilon(\mathbf{x}-\bm{\mathcal{X}^0}) + \frac{1}{2} \nabla \times \bm{\mathcal{N}} \psi_\epsilon(\mathbf{x}-\bm{\mathcal{X}^0})$ for a total force $\bm{\mathcal{F}}$ and total torque $\bm{\mathcal{N}}$ localized at $\bm{\mathcal{X}}^0$ \cite{Olson2013}. 
The blob function $\psi_\epsilon(\mathbf{x}-\bm{\mathcal{X}}^0)$ can be any smooth radial symmetric function approximating a three-dimensional Dirac delta distribution with the property that $\int_{0}^{\infty} \psi_\epsilon(r)dr=1$. Here, we use
\begin{equation}
\psi_\epsilon(r) = \frac{15\epsilon^4}{8\pi(r^2+\epsilon^2)^{7/2}}. \label{eq:blobfunction}
\end{equation}
where $r=|\mathbf{x}-\bm{\mathcal{X}}_0|$ is the radial distance from source point $\bm{\mathcal{X}_0}$. Details for the solutions of Stokes Eqs. (\ref{eq:stokes}) and (\ref{eq:cont}) for the velocity $\mathbf{u}^b(\mathbf{x})$ and angular velocity $\bm{\omega}^b(\mathbf{x})$ due to the described force $\mathbf{g}^b$ and blob function $\psi_\epsilon(r)$ can be found in \cite{Olson2013}; here we present the solution in matrix form as
\begin{eqnarray}
\mathbf{u}^b(\mathbf{x}) &=& \mathbf{S}^{\epsilon}(\mathbf{x}, \bm{\mathcal{X}}_0) \bm{\mathcal{F}} + \mathbf{Q}^{\epsilon}(\mathbf{x}, \bm{\mathcal{X}}_0) \bm{\mathcal{N}} ,\label{eq:stk1} \\   
\bm{\omega}^b(\mathbf{x}) &=& \mathbf{Q}^{\epsilon}(\mathbf{x}, \bm{\mathcal{X}}_0) \bm{\mathcal{F}} + \mathbf{D}^{\epsilon}(\mathbf{x}, \bm{\mathcal{X}}_0) \bm{\mathcal{N}} ,\label{eq:stk2}
\end{eqnarray} 
where $\mathbf{S}^{\epsilon}$ and $\mathbf{Q}^{\epsilon}$ are the regularized Stokeslet and regularized rotlet for the point force $\bm{\mathcal{F}}$ and torque $\bm{\mathcal{N}}$, respectively. The vorticity $\bm{\omega}^b(\mathbf{x})$ in Eq. (\ref{eq:stk2}) can be derived from Eqs. (\ref{eq:stk1}) by $\bm{\omega}^b=\frac{1}{2}\nabla \times \mathbf{u}$, and $\mathbf{D}^{\epsilon}$ is a regularized potential dipole.

For the hydrodynamic interactions of segments, in the centerline distribution of regularized Stokeslets, we assume that each segment includes just one regularized force and torque at its center position $\mathbf{X}_{m+1/2}$ \cite{Olson2013}. 
The forces and torques on segments ($\{\mathbf{F}_{m+1/2},\mathbf{N}_{m+1/2}\}$) can be calculated from the Kirchhoff rod theory as described in \S \ref{sec:Kirchhoff} from Eqs. (\ref{eq:torquebalance_dis}) and (\ref{eq:forcebalance_dis}).  Once the torques and forces are known, we calculate the local linear velocities $\mathbf{u}$ and the angular velocities $\bm{\omega}$ at the center of segments $\mathbf{X}_{m+1/2}$ by summing contributions like Eqs. (\ref{eq:stk1}) and (\ref{eq:stk2}) from all segments,
\begin{eqnarray}\label{eq:cnt-noslip}
\mathbf{u}(\mathbf{X}_{m+1/2}) &=& \sum_{k=0}^{M-1} \mathbf{S}^\epsilon_{k+1/2} \mathbf{F}_{k+1/2} + \sum_{k=0}^{M-1} \mathbf{Q}^\epsilon_{k+1/2} \mathbf{N}_{k+1/2},  \label{eq:cntu} \\
\bm{\omega}(\mathbf{X}_{m+1/2}) &=& \sum_{k=0}^{M-1} \mathbf{Q}^\epsilon_{k+1/2} \mathbf{F}_{k+1/2} + \sum_{k=0}^{M-1} \mathbf{D}^\epsilon_{k+1/2} \mathbf{N}_{k+1/2}.  \label{eq:cntw}
\end{eqnarray} 

By the no-slip boundary condition, these velocities are equal to the translational and angular velocities of segment $m+1/2$.
After finding the translational and rotational velocities, we update the new position and orientation of the segments using the forward Euler method. We use the superscript $n_t$ for time-step indexing such that time is $t = n_t \delta t$ and the associated position vector is $\mathbf{X}_{m+1/2}^{n_t}$, so 
\begin{eqnarray}
\mathbf{X}^{n_t+1}_{m+1/2} &=& \mathbf{X}^{n_t}_{m+1/2} + \mathbf{V}_{m+1/2}\delta t,\label{eq:update_pos} \\   
(\mathbf{d^i}_{m+1/2})^{n_t+1} &=& \left(\cos \theta \mathbf{I} + (1-\cos \theta)\mathbf{e}\mathbf{e}^T + \sin \theta (\mathbf{e}\times)\right)\left(\mathbf{d^i}_{m+1/2}\right)^{n_t}.   \label{eq:update_ori}
\end{eqnarray}
where $I$ is the $3\times3$ identity matrix, $\theta=|\bm{\Omega}_{p+1/2}|\delta t, \mathbf{e}=\bm{\Omega}_{p+1/2}/|\bm{\Omega}_{p+1/2}|$ and $(\mathbf{e}\times)$ is a matrix such that $(\mathbf{e}\times) \bm{v} = \mathbf{e}\times \bm{v}$ for any vector $\bm{v}$.

We also need to update the orthonormal triads at integer points $s_m$ used to calculate cross sectional transmitted forces and torques. To obtain the new orientations at integer points, we interpolate from the triads of segment centers $s_{m-1/2}$ and $s_{m+1/2}$ by (\cite{Olson2013,Lim2008}),
\begin{equation}
\mathbf{d}_{m}^i = \sqrt{\bm{A}} \mathbf{d}_{m-1/2}^i. \label{eq:update_di}
\end{equation}
where $i = 1,2,3$, $\bm{A}$ is a matrix defined by $\bm{A} = \sum_{i=1}^{3} {\mathbf{d}_{m-1/2}^i (\mathbf{d}_{m+1/2}^i)^T}$, and $\sqrt{\bm{A}}$ is the square root of matrix $\bm{A}$.

\begin{figure}[t]
	\centering
	\includegraphics[width=0.6\linewidth]{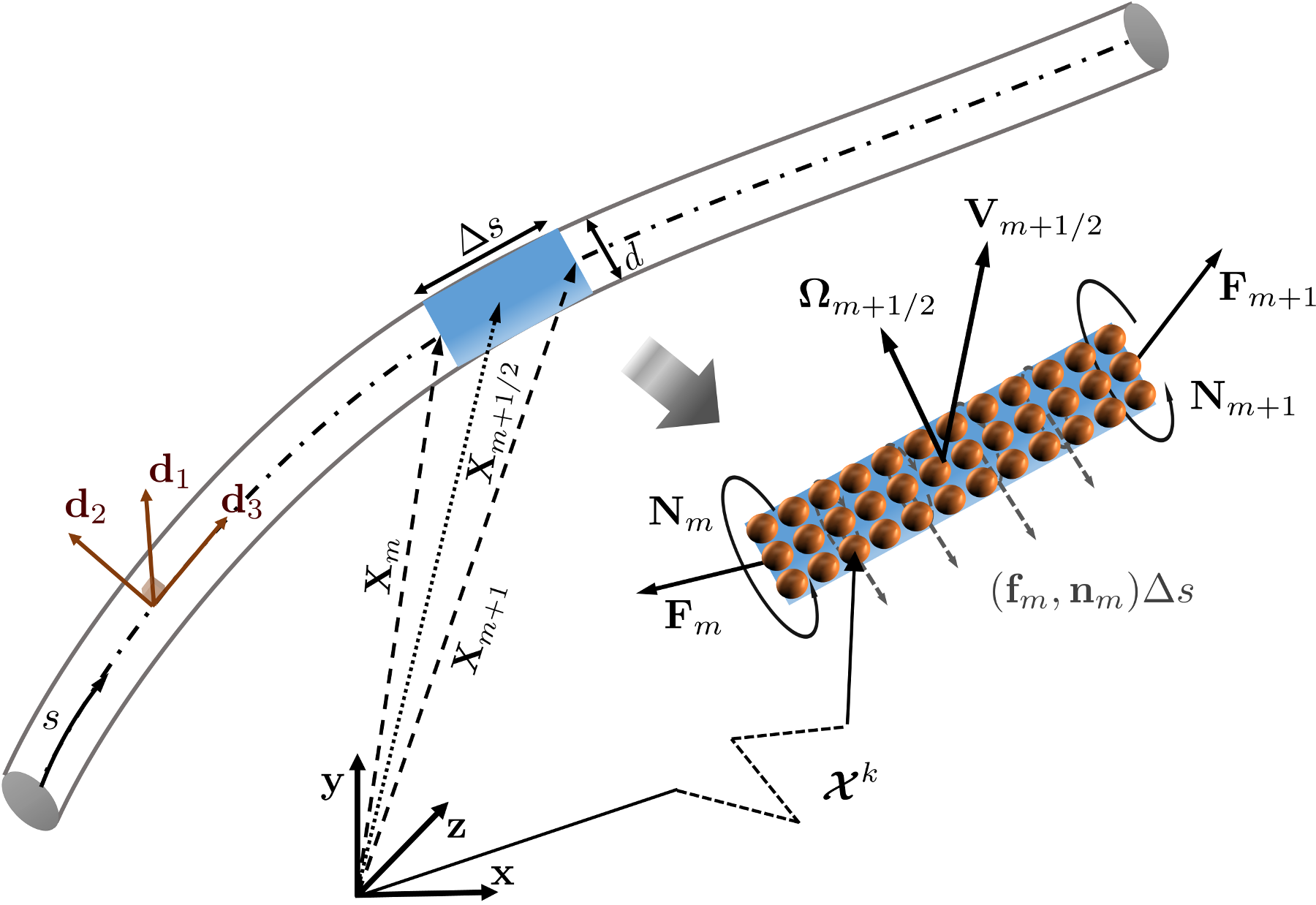}
	\caption{Surface discretization of filament by regularzied Stokeslets. The filament is discretized by cylindrical rigid-body segments of length $\Delta s$ and diameter $d$. We assume $N$ regularized Stokeslets are distributed on the surface of each  segment as shown by spheres at positions $\bm{\mathcal{X}}^k$ on the $m^{th}$ segment. The internal force and torque transmitted between cross sections specified by integer indices $m$ are calculated from the constitutive equation of the Kirchhoff rod model. Half-integer values $m+1/2$ label the centers of segments, and the translational and rotational velocities $\{ \mathbf{V}_{m+1/2},\bm{\Omega}_{m+1/2} \}$ are calculated to update the deformed position and orientation of the segments.}
	\label{fig:rodnum}
\end{figure}

\subsubsection{Method 2: Extensible Kirchhoff rod and surface distribution of regularized Stokeslets } \label{sec:ext}
In the second method we again assume that external forces and moments are found using the extensible Kirchhoff rod model through Eq. \ref{eq:torquebalance_dis} and \ref{eq:forcebalance_dis}.  However, we use a different, more accurate discretization of the filament by regularized Stokeslets.  In the centerline distribution of Method 1  (\S \ref{sec:cnt}), each segment contained just one regularized Stokeslet at its center.  Using the segment center point in Eq. \ref{eq:cnt-noslip}, the boundary conditions at the surface of segments are not completely satisfied. 
In this section, we describe how to use a distribution of regularized Stokeslets at the surface of segments to satisfy boundary conditions more accurately.

We discretize the surface of each segment with $N$ regularized Stokeslets (for a total number $MN$) with forces $\bm{\mathcal{F}}^k$ at collocation points ${\bm{\mathcal{X}}^k}$ (Fig. \ref{fig:rodnum}).  Then the torque at segment center $\mathbf{X}_{m+1/2}$ can be naturally calculated by $\sum_{k= N m +1}^{N(m+1)} (\mathbf{X}_{m+1/2}-\bm{\mathcal{X}}^k)\times \bm{\mathcal{F}}^k$. Therefore, we do not need to assign individual torques $\bm{\mathcal N}$ in Eq. (\ref{eq:stk1}) for the surface distributions of regularized Stokeslets, and the solution to Stokes equation only contains $\bm{\mathcal{F}}^k$ and simplifies to $\mathbf{u}^b=\mathbf{S}^\epsilon \bm{\mathcal{F}}$. 
For the distribution of regularized Stokeslets over all $M$ segments, using the linearity of Stokes equation, the velocity field at any point $\mathbf{x}$ can be written as,
\begin{equation}
\mathbf{u}(\mathbf{x}) = \sum_{k=1}^{MN} \mathbf{S}^\epsilon (\mathbf{x},\bm{\mathcal{X}}^k) \bm{\mathcal{F}}^k.  \label{eq:stk3}
\end{equation}
Evaluating Eq. (\ref{eq:stk3}) for the velocity $\mathbf{u}^k= \mathbf{u}(\bm{\mathcal{X}}^k)$ at the collocation points $\mathbf{x}=\bm{\mathcal{X}}^k$ can be described in a matrix form as, 
\begin{equation}
\begin{pmatrix}
\mathbf{u}^1 \\
\vdots	\\
\mathbf{u}^{MN}
\end{pmatrix} = 
\mathbf{G} 
\begin{pmatrix}
\bm{\mathcal{F}}^1 \\
\vdots	\\
\bm{\mathcal{F}}^{MN}
\end{pmatrix}.
\label{fv}
\end{equation}
where $\mathbf{G}\in \mathbb{R}^{3MN\times 3MN}$ is made of 3x3 block matrices; the block matrix at the $j^{th}$ row and $k^{th}$ column is $\mathbf{S}^\epsilon(\mathbf{x}_j,\mathbf{x}_k)$. 
Assuming rigid body motion for each segment, the total forces and torques of the $m^{th}$ segment can be described by point forces at the surface of that segment,
\begin{equation}
\begin{pmatrix}
\mathbf{F}_{m+1/2} \\
\mathbf{N}_{m+1/2}
\end{pmatrix} =  
\sum_{k= N m +1}^{N(m+1)}
\begin{pmatrix}
1&0&0 \\
0&1&0 \\
0&0&1 \\
0&-\Delta r^k_z&\Delta r^k_y\\
\Delta r^k_z&0&-\Delta r^k_x \\
-\Delta r^k_y&\Delta r^k_x&0
\end{pmatrix}
\bm{\mathcal{F}}^{k} \equiv 
\mathbf{K}^{\rm {T}}_{m+1/2} \begin{pmatrix}
\bm{\mathcal{F}}^1 \\
\vdots	\\
\bm{\mathcal{F}}^N
\end{pmatrix} \label{eq:K_f}
\end{equation}
where the last equality defines $\mathbf{K}\in \mathbb{R}^{3N\times 6}$, $\Delta \mathbf{r}^k=\bm{\mathcal{X}}^k-\mathbf{X}_{m+1/2}$ is the location of collocation point $k$ with respect to center of segment $m+1/2$,  and  $\mathbf{K}^{\rm T}$ is the transpose of the matrix $\mathbf{K}$ . 
Since the velocity of any collocation point $\mathbf{u}^k$ can be expressed by the translational and rotational velocities at the center of that segment by $\mathbf{u}^k=\mathbf{V}_{m+1/2}+\bm{\Omega}_{m+1/2}\times \Delta \mathbf{r}^k$, we can use the same matrix $\mathbf{K}_{m+1/2}$ to write \cite{Martindale2016}, 
\begin{equation}
\begin{pmatrix}
\mathbf{u}^{N m+1} \\
\vdots	\\
\mathbf{u}^{N (m+1)}
\end{pmatrix} = 
\mathbf{K}_{m+1/2} 
\begin{pmatrix}
\mathbf{V}_{m+1/2} \\
\bm{\Omega}_{m+1/2}
\end{pmatrix} \label{eq:K_u}
\end{equation}
Considering all collocation points and velocities at the centers of  segments,
\begin{equation}
\begin{pmatrix}
\mathbf{F}_{1/2} \\
\mathbf{N}_{1/2}\\
\vdots	\\
\mathbf{F}_{m+1/2} \\
\mathbf{N}_{m+1/2}\\
\vdots
\end{pmatrix}
= \mathbf{K}^{\rm T} \mathbf{G}^{-1} \mathbf{K} 
\begin{pmatrix}
\mathbf{V}_{1/2} \\
\bm{\Omega}_{1/2}\\
\vdots	\\
\mathbf{V}_{m+1/2} \\
\bm{\Omega}_{m+1/2}\\
\vdots
\end{pmatrix} \label{eq:HydroInteractions}
\end{equation}
where $\mathbf{K}\in \mathbb{R}^{3MN\times 6M}$ contains submatrices $\mathbf{K}_{m+1/2}$  and relates velocities of the collocations points to their corresponding centers' velocities and rotation rates \cite{Martindale2016}. 
The left hand side of Eq. (\ref{eq:HydroInteractions}) can be calculated by Eqs. (\ref{eq:forcebalance_dis}), and then we solve the system of equations in \ref{eq:HydroInteractions} for translational and angular velocities ($\{\mathbf{V}_{m+1/2}, \bm{\Omega}_{m+1/2}\}$) . After finding the translational and rotational velocities, we update the new position and orientation of segments using Eqs. (\ref{eq:update_pos}) and (\ref{eq:update_ori}).

\subsubsection{Method 3: Inextensible Kirchhoff rod and surface distribution of regularized Stokeslets } \label{sec:inx}
In the extensible version of the rod model used before for either centerline or surface distributions of regularized Stokeslets, a flexible filament can stretch and change its length during interactions with viscous flow. Here, we develop a new approach to model filament dynamics by enforcing inextensibility conditions. 
In the inextensible version of the Kirchhoff rod model, we enforce that the tangent vector of the centerline is aligned with $\mathbf{d}^3(s)$ and the centerline of filament is inextensible. So,
\begin{equation}
\frac{\partial \mathbf{X}}{\partial s} = \mathbf{d}^3,  \label{eq:inxcondX}
\end{equation} 
which implies that the norm $|\frac{\partial \mathbf{X}}{\partial s}| = 1$. The inextensibility constraints on velocities can be derived from Eq. (\ref{eq:inxcondX}) by differentiation with respect to time,  
\begin{equation}
\frac{\partial \mathbf{V}}{\partial s} = \bm{\Omega}(s)\times\mathbf{d}^3(s).  \label{eq:inxcondV}
\end{equation} 
where $\mathbf{V}(s)$ and $\bm{\Omega}(s)$ are the translational and rotational velocities along the centerline, respectively. 
For a filament discretized as described in \S \ref{sec:ext}, the discrete form of Eq. (\ref{eq:inxcondV}) is, 
\begin{eqnarray}
\mathbf{V}_{m} &=& \mathbf{V}_{m-1} + \bm{\Omega}_{m-1/2} \times (\mathbf{X}_m - \mathbf{X}_{m-1}),\label{eq:Vc_inx} \\   
\mathbf{V}_{m+1/2} &=& \mathbf{V}_{m} + \bm{\Omega}_{m+1/2} \times (\mathbf{X}_{m+1/2} - \mathbf{X}_{m}). \label{eq:Vm_inx}
\end{eqnarray}
These equations relate velocities at centers of segments ($\mathbf{V}_{m+1/2}$) and cross-sections ($\mathbf{V}_m$) to the rotational velocities at the center of segments ($\bm{\Omega}_{m+1/2}$). The constraints of inextensibility reduce the degrees of freedom from $6M$ (components of the translational and rotational velocities of segments) for the extensible model to $3+3M$ ($3$ components of filament's overall translational velocity and $3M$ components of the rotational velocities of segments). The matrix form of Eqs. (\ref{eq:Vc_inx}) and (\ref{eq:Vm_inx}) are useful and can be written as,

\begin{equation}
\begin{pmatrix}
\mathbf{V}_{1/2} \\
\bm{\Omega}_{1/2}\\
\vdots	\\
\mathbf{V}_{M+1/2} \\
\bm{\Omega}_{M+1/2}\\
\end{pmatrix}
= \mathbf{L} 
\begin{pmatrix}
\mathbf{V}_{1/2} \\
\bm{\Omega}_{1/2}\\
\bm{\Omega}_{3/2}\\
\vdots \\
\bm{\Omega}_{M+1/2}\\
\end{pmatrix} \label{L_inx}
\end{equation}
where $\mathbf{L}\in \mathbb{R}^{6M\times3(M+1)}$ calculates the segments' velocities from their corresponding rotational velocities using Eqs. (\ref{eq:Vc_inx}) and (\ref{eq:Vm_inx}). 

Corresponding to the reduction of kinematic degrees of freedom is a reduction in the number of forces and torques needed to specify the motion.  As we show below it is sufficient to only specify the internal torques.  Assuming we have a known configuration, the cross-sectional torques $\mathbf{N}_m$ are given by Eq. \ref{eq:torquebalance_dis}.  For the inextensible model, the constitutive model for forces (Eq. \ref{eq:rodforce} and \ref{eq:forcebalance_dis}) does not apply; rather the internal forces are determined by whatever is needed to satisfy the inextensibility constraint.
The force and torque at the free end of the filaments is zero $(\mathbf{F}_{M}=\mathbf{0},\mathbf{N}_{M}=\mathbf{0})$. Thus, from Eq. \ref{eq:forcebalance_dis} we can express the internal transmitted forces at the $m^{th}$ cross section point as summation of fluid forces acting on segments from that cross section  to the free end of the filament,
\begin{equation}
\mathbf{F}_m = \sum_{i=m}^{M-1} {\mathbf{F}_{i+1/2}}. \label{eq:force_inx}
\end{equation}
We rewrite Eq. (\ref{eq:torquebalance_dis}) such that the cross sectional torques can be described as a function of force and torques of segments as,
\begin{equation}
\mathbf{N}_{m} - \mathbf{N}_{m+1} = \frac{\Delta s}{2}(\mathbf{d}^3_{m+1}\times \mathbf{F}_{m+1} + \mathbf{d}^3_{m}\times \mathbf{F}_{m}) + \mathbf{N}_{m+1/2}.\label{eq:torquebalance_inx}
\end{equation}
Eqs. (\ref{eq:force_inx}) and (\ref{eq:torquebalance_inx}) are a set of linear equations that can be summarized in matrix form as
\begin{equation}
\begin{pmatrix}
\mathbf{F}_0 \\
\mathbf{N}_0 \\
\mathbf{N}_1 \\
\vdots	\\
\mathbf{N}_{M-1} \\
\end{pmatrix}
= \mathbf{J} 
\begin{pmatrix}
\mathbf{F}_{1/2} \\
\mathbf{N}_{1/2}\\
\vdots \\
\mathbf{F}_{M + 1/2} \\
\mathbf{N}_{M + 1/2}
\end{pmatrix} \label{eq:J_inx},
\end{equation}
where $\mathbf{J}\in \mathbb{R}^{3(M+1)\times6M}$. 
The hydrodynamic interactions are calculated as previously described in \S\ref{sec:ext} for the surface distribution of regularized Stokeslets using Eqs. (\ref{eq:stk3}), (\ref{fv}) and (\ref{eq:HydroInteractions}).	
Combining Eqs. (\ref{L_inx}) , (\ref{eq:J_inx}), and (\ref{eq:HydroInteractions}), we have 
\begin{equation}
\begin{pmatrix}
\mathbf{F}_0 \\
\mathbf{N}_0 \\
\vdots	\\
\mathbf{N}_{M-1} \\
\end{pmatrix}
= \mathbf{J} (\mathbf{K}^T \mathbf{G}^{-1} \mathbf{K}) \mathbf{L}
\begin{pmatrix}
\mathbf{V}_{1/2} \\
\bm{\Omega}_{1/2}\\
\vdots \\
\bm{\Omega}_{M + 1/2}
\end{pmatrix}. \label{eq:inx}
\end{equation}
We solve Eq. (\ref{eq:inx}) to find the translational velocity of the filament $\mathbf{V}_{1/2}$ and rotation rates at the center of segments $\bm{\Omega}_{m+1/2}$. Then, we update the positions and orientations of the segments by Eqs. (\ref{eq:update_pos}) and (\ref{eq:update_ori}). The position of integer points $m$ are calculated by updating with velocities from Eq. (\ref{eq:Vc_inx}) and corresponding orientations are obtained by \ref{eq:update_di}.

\subsection{Geometry, boundary conditions, and discretization } \label{sec:geometry}
As a test case, in this paper we simulate a helical bacterial flagellar filament rotating in a viscous flow. Initially the undeformed filament is a tapered helix along the $x$-direction with filament diameter $d=0.032 \rm \mu m$, helical radius $R=0.14 \rm \mu m$, and helical pitch $P=1.49 \rm \mu m$, and centerline described by
\begin{equation}
\mathbf{r}_c (l)=l \hat{\mathbf{x}} + R(1-e^{-(2\pi l/P)^2 })[\cos(2\pi l/P)\hat{\mathbf{y}} + \sin(2\pi l/P)\hat{\mathbf{z}}], \label{eq:flag}
\end{equation}
where  $l \in [0, 3.9692 \rm \mu m]$) so that the total contour length of the curve is $L=4.59 \rm \mu m$.

For the hydrodynamic interactions of the surface distribution of regularized Stokeslets, we use a uniform distribution of regularized Stokeslets on the surface of filament. We specify 12 Stokeslets on each cross-sectional circumference and each cross-section is spaced along the arclength by $h=\pi d/12$. Thus, Stokeslets have a typical spacing of $h$ and there are a total number of $6620$ Stokeslets. The blob parameter for the surface distributions was chosen based on Stokeslet separation as ${\epsilon} = h/3$ as described in \cite{Martindale2016}.  

For the centerline distribution approach, we assume that each segment has only one regularized stokeslet and rotlet. For this case, the blob parameter should be chosen to be close to the filaments' diameter. Thus, segment lengths are also limited to a range close to the diameter of filament. To find an appropriate blob parameter for the centerline distribution, we calculate the torque of a rigid helical filament with the geometry in Eq. (\ref{eq:flag}) rotating with a prescribed angular velocity using both the surface and centerline distributions, and choose the blob parameter for the centerline distribution so that they match, as described in \cite{Martindale2016}. We find that the optimal blob parameter for the centerline distributions of regularized Stokeslets (\textcolor{black}{$\epsilon_{cnt}$}) can be fit by $(\epsilon_{cnt}/d) = 0.187+0.561(\Delta s/d)$ with $\textcolor{black}{\rm{R\!\!-\!\!squared}} = 0.9985$.

In the Results we compare different numerical approaches \textcolor{black}{when the base of the filament is prescribed to rotate with angular velocity $\omega$, i.e. $\bm{\Omega}_{1/2} = (\omega,0,0)$, and the base of filament does not translate, i.e. $\bm{V}_{1/2} = (0,0,0)$}, and we solve for the shape and force/torques of the segments over time.  Solving Eqs. (\ref{eq:HydroInteractions}) and (\ref{eq:inx}) requires that $\{\mathbf{F}_{1/2}, \mathbf{N}_{1/2}\}$ or $\{\mathbf{F}_{0}, \mathbf{N}_{0}\}$, respectively, are treated as unknowns corresponding to the total torque and force required to generate the prescribed motion; all other forces and torques on the left hand side of those equations are specified by the filament geometry at each time-step.  Similarly in Eqs. (\ref{eq:cntu}) and (\ref{eq:cntw}), $\{ \mathbf{u}(\mathbf{x}_{1/2}) = \mathbf{V}_{1/2},  \bm{\Omega}(\mathbf{x}_{1/2}) = \bm{\Omega}_{1/2} \}$ are knowns while 
$\{\mathbf{F}_{1/2}, \mathbf{N}_{1/2}\}$ are unknowns.  
 \textcolor{black}{For ${EI}=10~ \rm pN \mu m^2$ and ${EI}=1~ \rm pN \mu m^2$, the non-dimensional prescribed rotation rates  $\bar{\Omega}=\textcolor{black}{\omega} \mu L^4/{EI}$ are 5 and 50, respectively,} well below the onset of instabilities in the dynamics at $\bar{\Omega} \sim 300$ \cite{Jawed2017,Park2017}.

To compare the deformed shape of a filament to another reference shape, we average the difference in positions between the deformed ($\mathbf{x}$) and reference ($\mathbf{x}_r$) shapes, and normalize by the filament helical radius $R$, 
\begin{eqnarray}
er = \frac{1}{M} \frac{\sum_{p=1}^{M} |\mathbf{x}^p-\mathbf{x}^p_{r}|}{R}. \label{eq:error}
\end{eqnarray}
where $|\mathbf{x}|$ is the norm of vector $\mathbf{x}$.  The positions are sampled at $M$ locations indexed by $p$, where $M$ is the number of segments in the more finely discretized filament.

\subsection{Partial updating of hydrodynamic interactions} \label{sec:hydro-update}
For the surface distribution of regularized Stokeslets, calculating the hydrodynamic interactions at each time-step is computationally expensive. Since the time-steps are small and controlled by the \textcolor{black}{elastic bending dynamics}, we can maintain accuracy without updating the hydrodynamic interactions at each time-step.  Here, we describe 
how we calculate filament kinematics without updating the hydrodynamic interactions and its effect on accuracy.

For a filament discretized \textcolor{black}{into} segments, the translational and rotational velocities of segments depend on their relative displacements and orientations with respect to each other. For a given configuration of a deformed filament at time $t$, we define a generalized configuration vector \textcolor{black}{$\bm{\mathscr{X}}(t)=\{\mathbf{X}_{m+1/2}(t),\mathbf{d}^i_{m+1/2}(t)\}$, for $m=0$ to $M\!-\!1$,} containing the positions and orientations of all segments. The internal force and torques can be calculated in a generalized vectorial form \textcolor{black}{$\bm{\mathscr{N}}(t) = \{\mathbf{F}_{m+1/2}(t),\mathbf{N}_{m+1/2}(t)\}$, for $m=0$ to $M\!-\!1$,} as a function of generalized configuration vector $\bm{\mathscr{X}}$ from the discrete form of constitutive relations (\ref{eq:rodforce_dis}) and ({\ref{eq:rodtorque_dis}) by $\bm{\mathscr{N}}(t) = \bm{\mathscr{F}}(\bm{\mathscr{X}}(t))$.
Then, we can evaluate the translational and rotational velocities via the hydrodynamics using Eq. (\ref{eq:HydroInteractions}). Thus, defining \textcolor{black}{ the mobility matrix} $\textcolor{black}{\mathscr{M}}=(\mathbf{K}^T \mathbf{G}^{-1} \mathbf{K})^{-1}$, we can say that the generalized vector of velocities \textcolor{black}{$\bm{\mathscr{V}}(t)=\{\mathbf{V}_{m+1/2}(t),\bm{\Omega}_{m+1/2}(t)\}$, for $m=0$ to $M\!-\!1$,} are related to the configuration vector $\bm{\mathscr{X}}(t)$ as $\bm{\mathscr{V}}(t) = \textcolor{black}{\mathscr{M}}(\bm{\mathscr{X}}(t))\bm{\mathscr{F}}(\bm{\mathscr{X}}(t))$. 
Due to the stiffness of the filament, forces and torques (hence $\bm{\mathscr{F}}(\bm{\mathscr{X}}(t))$) are much more sensitive to changes in configuration vector $\bm{\mathscr{X}}(t)$ than the hydrodynamic interactions ($\textcolor{black}{\mathscr{M}}$). To investigate this dependency, we Taylor expand in time $t+\delta t$ to express velocities as 
\begin{eqnarray}
\bm{\mathscr{V}}(t+\delta t) &=& \textcolor{black}{\mathscr{M}}\Big(\bm{\mathscr{X}}(t+\delta t)\Big)\bm{\mathscr{F}}\Big(\bm{\mathscr{X}}(t+\delta t)\Big)=\Big[\textcolor{black}{\mathscr{M}}+\frac{\partial \textcolor{black}{\mathscr{M}}}{\partial \bm{\mathscr{X}}}\Delta \bm{\mathscr{X}} + \cdots\Big]\bm{\mathscr{F}}\Big(\bm{\mathscr{X}}(t+\delta t)\Big)  \nonumber \\
&=& \textcolor{black}{\mathscr{M}}\Big(\bm{\mathscr{X}}(t)\Big) \bm{\mathscr{F}}\Big(\bm{\mathscr{X}}(t+\delta t)\Big) + \Big[\frac{\partial \textcolor{black}{\mathscr{M}}\Big(\bm{\mathscr{X}}(t)\Big)} {\partial \bm{\mathscr{X}}}\Delta \bm{\mathscr{X}} + \cdots\Big]\bm{\mathscr{F}}\Big(\bm{\mathscr{X}}(t+\delta t)\Big) \nonumber \\
&=& \bm{\mathscr{V}}_0 (t+\delta t) ~~~~~~~~+ \Delta \bm{\mathscr{V}} ,
\label{eq:hydro-update}
\end{eqnarray}
where $\bm{\mathscr{V}}_0$ is the approximate velocity used when hydrodynamic interactions are not updated.
Thus, the velocity difference $\Delta \bm{\mathscr{V}}$ is the error due to not updating hydrodynamic interactions over a time interval $\delta t$. If the error is too large, we need to update the hydrodynamic interactions by recalculating $\textcolor{black}{\mathscr{M}}$ for the new current configuration $\mathscr{X}(t+\delta t)$. In Fig. \ref{fig:hydro-update}, we examine the magnitude of errors $\Delta \bm{\mathscr{V}}$ averaged over all segments. 
The errors are generated by using the initial configuration Eq. (\ref{eq:flag}) and randomly perturbing the positions and orientations of all the segments; the quantity $|\Delta \mathscr{X}|$ is the maximum displacement of any segment in the filament from the initial configuration.
For the rest of this paper, we update the hydrodynamic interactions when the biggest relative displacement of the segments passes the criterion  $|\Delta \bm{\mathscr{X}}(t)|/d>0.2$ for which corresponding errors are expected to be less than $10^{-3}\%$.  
\begin{figure}[!t]
	\centering
	\includegraphics[width=0.45\linewidth]{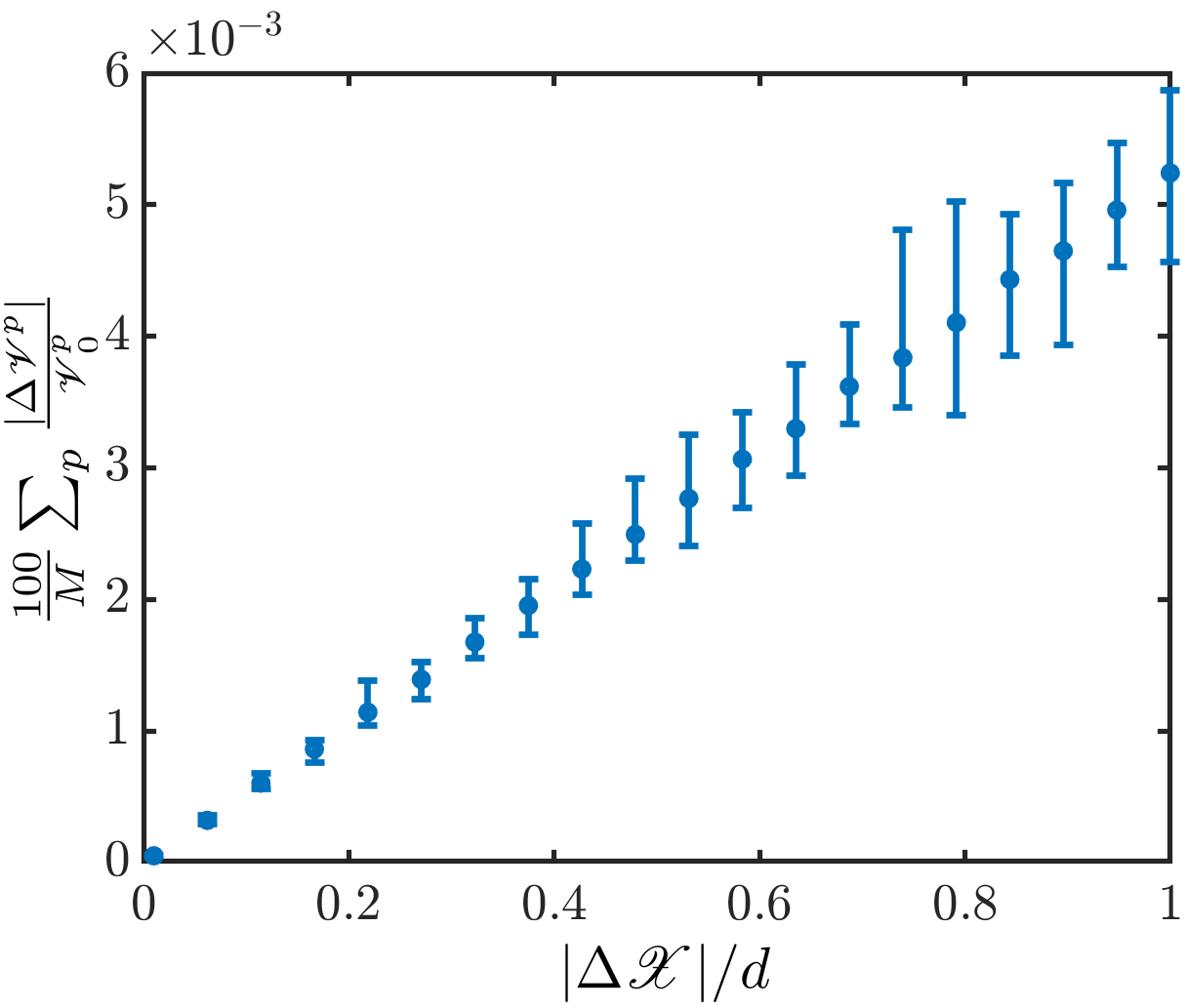}
	\caption{ The errors in computed velocities expected when hydrodynamic interactions are not updated for perturbations of configuration vector $\bm{\mathscr{X}}$ of magnitude $|\Delta \bm{\mathscr{X}} |$. }
	\label{fig:hydro-update}
\end{figure}  

\subsection{The effects of segment sizes on the accuracy of numerical approaches} \label{sec:segments}
We discretize our filament into straight segments of length $\Delta s$ as described in \S \ref{sec:ext}. The choice of segment size can affect the accuracy of numerical models as well as the bending and stretching timescales shown in Fig. \ref{fig:timescale}. 
To ensure an accurate representation of the geometry, the largest allowed $\Delta s$ is determined by the typical curvature of the filament arising from the pitch or taper of the helix or its dynamic deformation. For a standard helix, the curvature of  filament depends on the helical radius $R$ and pitch $p$ by $\kappa_0=R/(R^2+(p/2\pi)^2)$ while for the tapered helix used in this paper (Eq. (\ref{eq:flag})), the maximum curvature is at the base of filament $x=0$. For the parameters of our helical model, the standard and maximum curvature are $\kappa_0 \approx1.84 \rm 1/\mu m$ and $\kappa_{max}\approx 4.98 \rm 1/\mu m$, respectively, which give estimates for choosing suitable lengthscales for segment sizes $\Delta s$. 

In the results of this manuscript, we do not focus on the effect of varying segment size and keep the filament geometry and segment size constant.  However, we need to choose an appropriate segment size to obtain accurate results, which we have done as follows.
For a constant filament diameter, we investigate discretization sizes in the range of $\Delta s/d \in [ 0.75,7]$. The finest segmentation,  $(\Delta s/d)=0.75$, corresponds to the largest curvature as $\kappa_{max} \Delta s=0.1$ and the coarsest segmentation, $(\Delta s/d)=7$, corresponds to $\kappa_{max} \Delta s=1$.
We compare the effects of varying segment sizes for different methods 
by calculating the error in filament shape relative to the shape of the extensible model with the finest discretization, $(\Delta s/d)=0.75$.  Fig. \ref{fig:segments2} plots the maximum error over the time interval $0<\omega t<15$ as a function of segment size. 
For the rest of paper, we use $(\Delta s/d)=1$ in our numerical simulations which gives discretization errors  less than $0.5\%$. 


\begin{figure}[!t]
	\centering
	\includegraphics[width= 0.55 \linewidth]{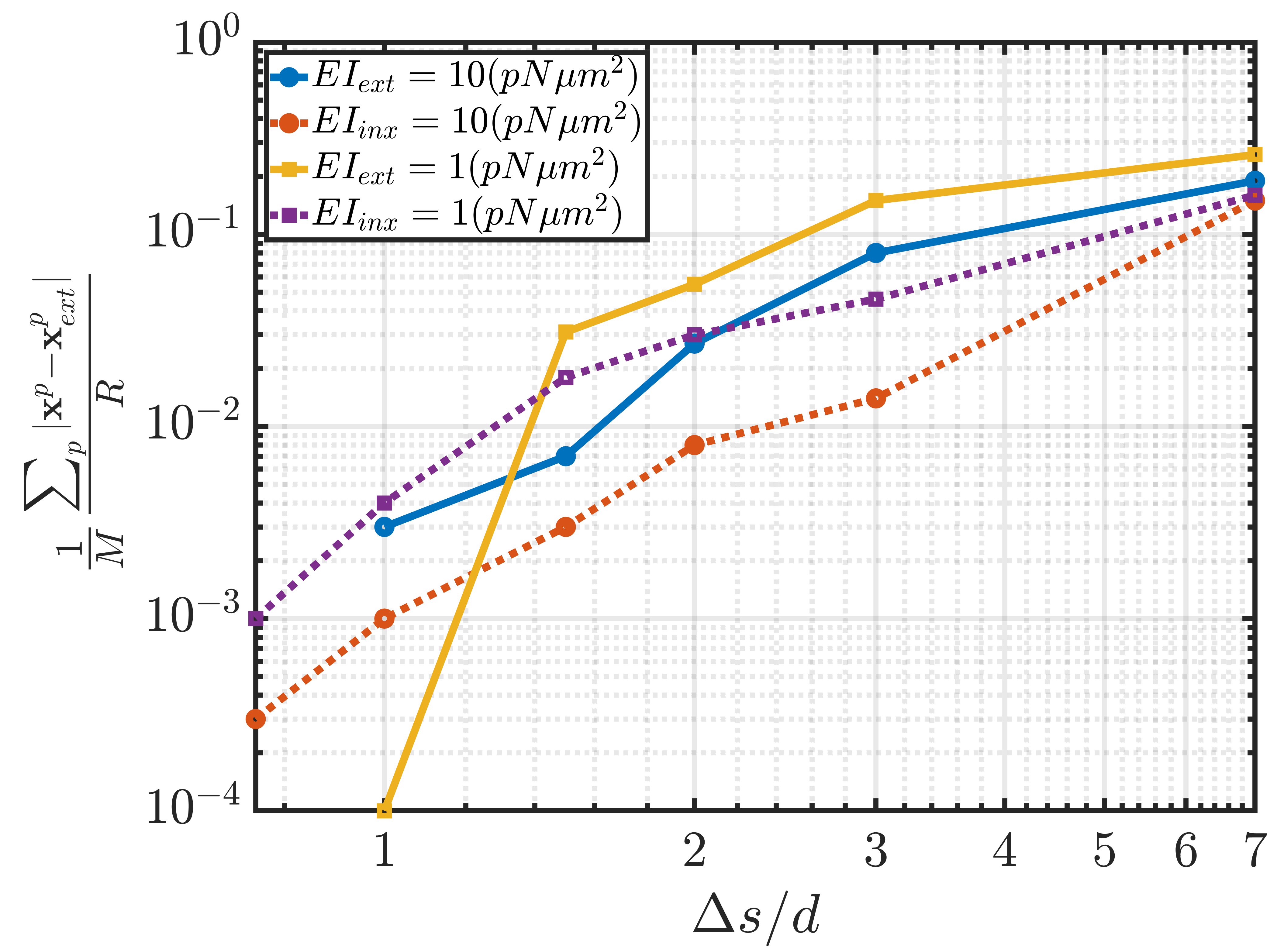}
	\caption{The maximum errors during the time interval $0<\omega t<15$ as a function of segment size for different numerical approaches and bending rigidities. The effective diameter in this plot is ${d_{\rm eff}}=50 \rm nm$ and the reference model is the extensible model with the finest discretization $(\Delta s/d=0.75)$.}
	\label{fig:segments2}
\end{figure}

\section{Results}

To assess the increase in computational efficiency associated with removing stretching timescales in the inextensible model, we compare the performance of the three methods for the different combinations of stretching and bending stiffnesses summarized in Table \ref{table:EAEI}.  In Fig. \ref{fig:timescale}, the bending and stretching timescales depend on $\Delta s/d$, since $\tau_s/\tau_b = {EI}/({EA}~\Delta s^2) \sim (\Delta s/d)^{-2}$.  However, in all of our tests, we use a fixed filament geometry with diameter $d=0.032 \,\mu$m \textcolor{black}{(so that $L/d>100$ and the Kirchoff rod approximation is always valid)} and a fixed segment size $\Delta s = d$ as described in  \S\ref{sec:segments} so that discretization errors are well-controlled.  Instead, we vary the ratio of stretching and bending timescales by treating ${EI}$ and ${EA}$ in our model as independent parameters.  This can be interpreted as an effective diameter that sets the ratio of timescales; assuming a circular cross section for filaments so that $I = \pi d^4/64$ and $A = \pi d^2/4$, the effective diameter is ${d_{\rm eff}} = 4/\sqrt{{EA/EI}}$.  The values of ${d_{\rm eff}}$ for our trials are reported in Table \ref{table:EAEI} and cover a wide range of typical biological filament sizes in viscous flow.

\begin{table}[!t]
	\begingroup
	\setlength{\tabcolsep}{6pt}
	\renewcommand{\arraystretch}{1}

	\centering
	\caption{Different combinations of the stretching and bending stiffnesses with associated timescales and effective diameters that are studied in this paper.}
	
	\begin{tabular}{|cc|c@{}|c|c|c|} 
		\hline
		$\mathbf{EA/EI~\rm (1/\mu m^2)}$ &  & $\mathbf{100}$ & $\mathbf{1000}$ & $\mathbf{10,000}$ & $\mathbf{100,000}$ \\ \hline
		$\mathbf{d_{eff}~\rm (nm)}$ &   & $\mathbf{200}$ & $\mathbf{100}$ & $\mathbf{50}$ & $\mathbf{10}$ \\ \hline
		\multicolumn{1}{|c|}{\multirow{2}{*}{$\mathbf{EI~ \rm (pN \bm{\mu} m^2)}$}} & $\mathbf{1}$ & \begin{tabular}[c]{c}$\tau_b \sim 10^{-8}$\\ $\tau_s \sim 10^{-7}$\\ $\tau_s /\tau_b \sim 10$\end{tabular} & \begin{tabular}[c]{c}$\tau_b \sim 10^{-8}$\\ $\tau_s \sim 10^{-8}$\\ $\tau_s /\tau_b \sim 1$\end{tabular} & \begin{tabular}[c]{c}$\tau_b \sim 10^{-8}$\\ $\tau_s \sim 10^{-9}$\\ $\tau_s /\tau_b \sim 0.1$\end{tabular} & \begin{tabular}[c]{c}$\tau_b \sim 10^{-8}$\\ $\tau_s \sim 10^{-10}$\\ $\tau_s /\tau_b \sim 0.01$\end{tabular} \\ \cline{2-6} 
		\multicolumn{1}{|c|}{} & $\mathbf{10}$ & \begin{tabular}[c]{c}$\tau_b \sim 10^{-9}$\\ $\tau_s \sim 10^{-8}$\\ $\tau_s /\tau_b \sim 10$\end{tabular} & \begin{tabular}[c]{c}$\tau_b \sim 10^{-9}$\\ $\tau_s \sim 10^{-9}$\\ $\tau_s /\tau_b \sim 1$\end{tabular} & \begin{tabular}[c]{c}$\tau_b \sim 10^{-9}$\\ $\tau_s \sim 10^{-10}$\\ $\tau_s /\tau_b \sim 0.1$\end{tabular} & \begin{tabular}[c]{c}$\tau_b \sim 10^{-9}$\\ $\tau_s \sim 10^{-11}$\\ $\tau_s /\tau_b \sim 0.01$\end{tabular} \\ \hline
	\end{tabular}
	
	\label{table:EAEI}
	\endgroup
\end{table}
\subsection{Time-step convergence study for the different approaches} \label{sec:time-conv}
Previously, we estimated stretching and bending timescales as a function of filament stiffness and segmentation sizes in \S \ref{sec:timescale} and Fig. \ref{fig:timescale}. We expect that the smallest value of these timescales (bending $\tau_b$ , or stretching $\tau_s$) controls the necessary time-step  in numerical solutions. To test this, we first set $\delta t_0=\min\{\tau_s,\tau_b\}$ as the reference time-step size for the extensible models, and $\delta t_0 = \tau_b$ as the reference time-step for the inextensible model.  Then we check the time step convergence by choosing time-steps smaller than $\delta t_0$ and comparing the maximum errors in the shape (Eq. (\ref{eq:error})) over the time interval $0<\omega t<15$ relative to the shape for the smallest timestep, $\delta t= 0.01 \delta t_0$.  Fig. \ref{fig:timeconv} shows 
small errors (less than $5\times 10^{-4}\%$) for all choices of time-steps, implying that the estimated reference time-steps are suitable for the inextensible and extensible models.  All other calculations reported use the reference time-steps.

\begin{figure}[!t]
	\centering
	\includegraphics[width= 0.5\linewidth]{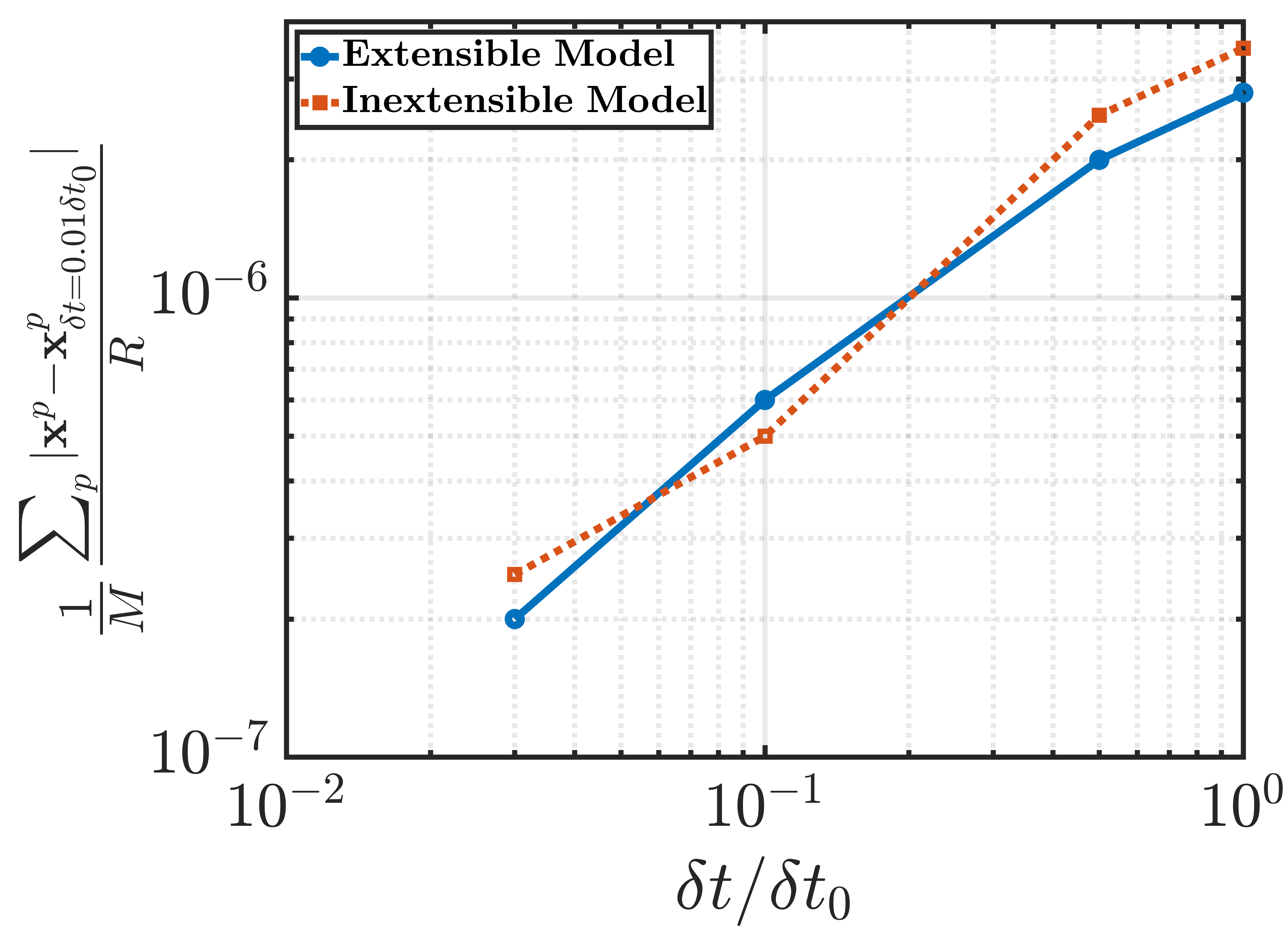}
	\caption{Timestep convergence of numerical approaches for extensible and inextensible models. 
		For these test cases the bending and stretching rigidity are ${EI}=10~\rm pN \mu m^2$ and ${EA/EI} = 10^4~ \rm 1/\mu m^2$, respectively and the segment sizes are $\Delta s/{d} = 1$.  For the extensible model, the reference time-step $\delta t_0$ is chosen from Table \ref{table:EAEI} such that $\delta t_0=\min \{\tau_s,\tau_b\}$ for a given combination of $\{{EI,EA/EI}\}$; for this test case $\delta t _0=10^{-12}s$. For the inextensible model the reference time scale is $\delta t_0=\tau_b=10^{-9}s$. We compare maximum errors in deformed shapes of filaments during the time $0<\omega t<15$ relative to the shape of the filament for the smallest timestep $\delta t = 0.01 \delta t_0$.}
	\label{fig:timeconv}
\end{figure}

\subsection{Extensibility of filaments} \label{sec:extensibility}

In the next two sections we evaluate the accuracy of the inextensible model.  First, we investigate how much the lengths of extensible filaments change for varying values of stretching and bending stiffnesses.
We measure the change in the total length of the filament compared to its initial arclength $L_0$. For the inextensible model, the change in total length is zero because of the inextensibility condition (Eq.  (\ref{eq:inxcondX})). For the extensible model, the change in length between the $m$ and $m+1$th segments is 
\begin{equation}
\Delta l_m = [(\mathbf{X}_{m+1/2}-\frac{\Delta s}{2}\mathbf{d}^3_{m+1/2})-(\mathbf{X}_{m-1/2}+\frac{\Delta s}{2}\mathbf{d}^3_{m-1/2})]\cdot \mathbf{d}^3_m.
\label{eq:stretch}
\end{equation}
The total change in the arclength is the summation of Eq. (\ref{eq:stretch}) over all $M$ segments, $\Delta l = \sum_{k=0}^{M-1} {\Delta l_k}$. The total length over time can be expressed as $L(t) = M \Delta s + \Delta l$.

The results are plotted in Fig. \ref{fig:extensibility} for bending rigidities of ${EI} = 10~\rm pN \mu m^2$ and $1~\rm pN \mu m^2$ with different effective filament diameter ${d_{\rm eff}}$. This figure shows that filaments are almost inextensible with changes in length that increase as ${d_{\rm eff}}$ increases (i.e., as the stretching modulus ${EA}$ decreases).
The maximum change in the length is about $0.2\%$ for filaments with effective diameter of ${d_{\rm eff}}=200 \rm nm$. The method of centerline distribution of regularized Stokeslets predicts more extensibility for filaments compared to surface distribution of regularized Stokeslets.

\begin{figure}[!t]
	\centering
	\includegraphics[width=\linewidth]{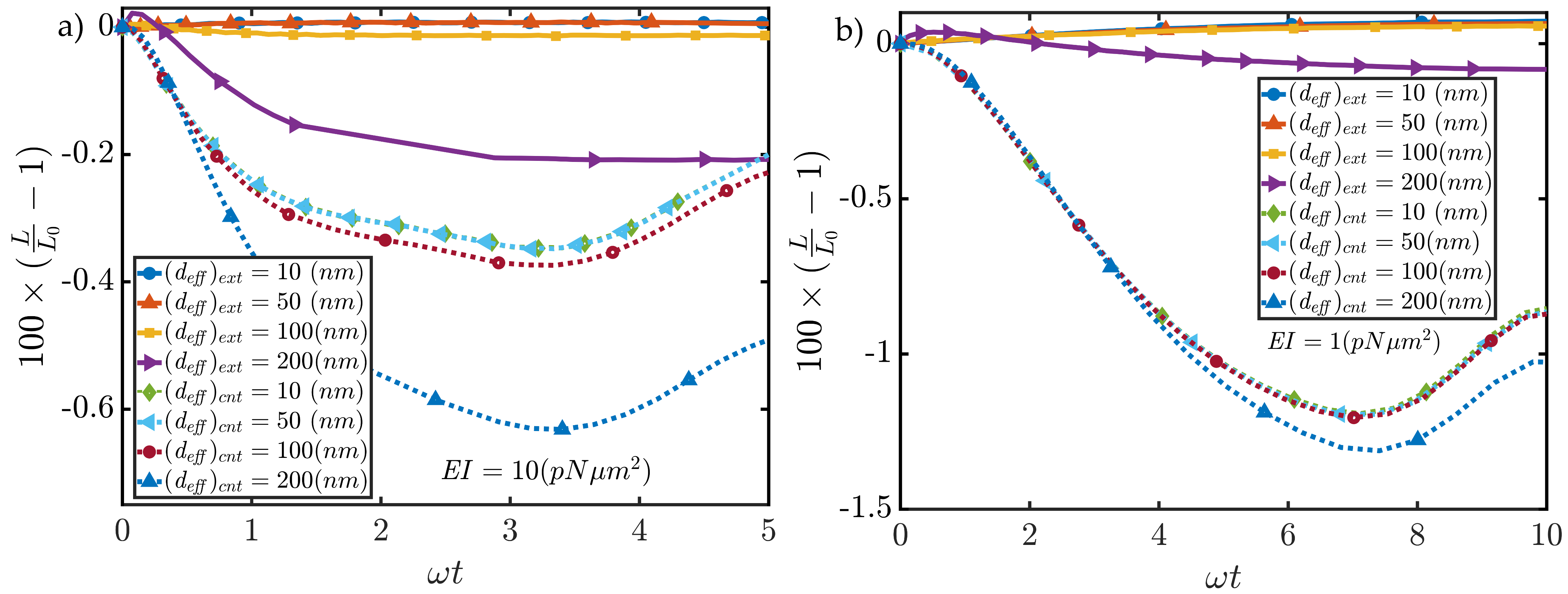}
	\caption{Extensibility of the filaments with different bending and stretching rigidities over time as calculated by Eq. (\ref{eq:stretch}). For the bending rigidity of a) $EI = 10~ \rm pN \mu m^2$ and b) $EI =~1 \rm pN \mu m^2$. The maximum extensibility is $0.2\%$ for effective filament diameter ${d_{\rm eff}} = 200 \rm nm$. The method of centerline distribution of regularized Stokeslets (dotted lines) predicts more stretching for the filaments compared to the surface distribution of regularized Stokeslets (solid lines).}
	\label{fig:extensibility}
\end{figure}

\subsection{Accuracy of the inextensible model} \label{sec:accuracy}
In \S \ref{sec:extensibility}, we show that the filaments are almost inextensible for different stretching and bending rigidities and \textcolor{black}{small} effective diameters. Thus, we expect that one can use the inextensible model to describe filaments bending dynamics. In this section, we quantify the accuracy of the inextensible model by comparing it to the extensible approaches. In Fig. \ref{fig:shape}, we plot the errors (Eq. (\ref{eq:error})) in the deformed shapes of a filament with various stretching/bending ratios modeled by the different numerical approaches. We present results for two bending rigidities of ${EI} = 1~ \rm pN \mu m^2$ and $10~ \rm pN \mu m^2$.  The reference case for the error is the inextensible filament for each value of ${EI}$. Our results show that as long as ${d_{\rm eff}}<50 \rm nm$ the inextensible model differs from the extensible model with surface distribution of regularized Stokeslets by less than $1\%$ of the helical radius for both values of $EI$. 
The centerline distribution of regularized Stokeslets leads to the largest differences in shape from both the extensible and inextensible models, suggesting that it is significantly less accurate.
We also plot the maximum steady state error between the inextensible and extensible models for the two values of ${EI}$ in 
Fig. \ref{fig:inxshape}.  For bending rigidities with ${d_{\rm eff}}<50 \rm nm$ the errors in the steady state shapes are less than $1\%$ of the helical radius $R$. 

\begin{figure}[!t]
	\centering
	\includegraphics[width=\linewidth]{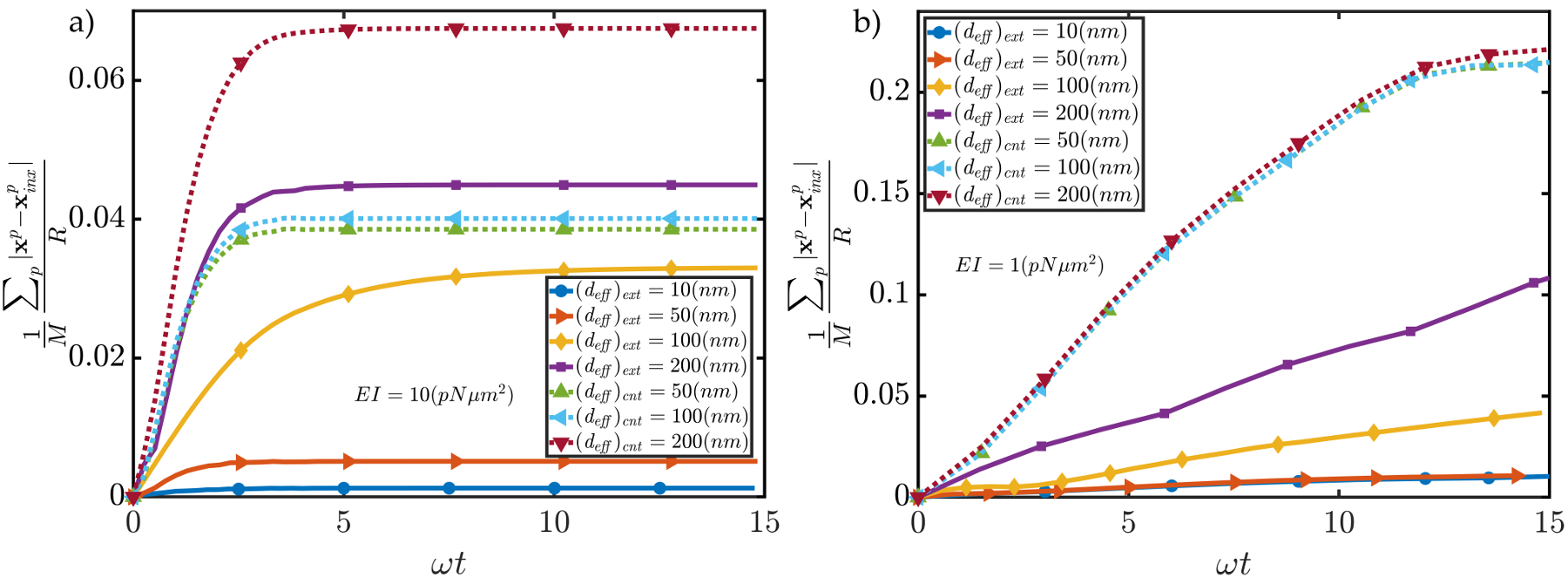}
	\caption{Comparing the deformed shapes of filaments with different effective filament diameter and bending rigidities, for a) $EI = 10~ \rm pN \mu m^2$, and b) $EI = 1~ \rm pN \mu m^2$. For each effective filament diameter ${d_{\rm eff}}$ the shapes from various numerical models are compared to the reference inextensible model. The extensible model predicts shapes with errors less than $1\%$ compared to the inextensible models for ${d_{\rm eff}}<50 \rm nm$, while for ${d_{\rm eff}}<100 \rm nm$ errors are less than $5\%$. The centerline models are much less accurate especially for the less stiff filaments with $EI = 1~ \rm pN \mu m^2$.}
	\label{fig:shape}
\end{figure}

\begin{figure}[h]
	\centering
	\includegraphics[width= 0.8\linewidth]{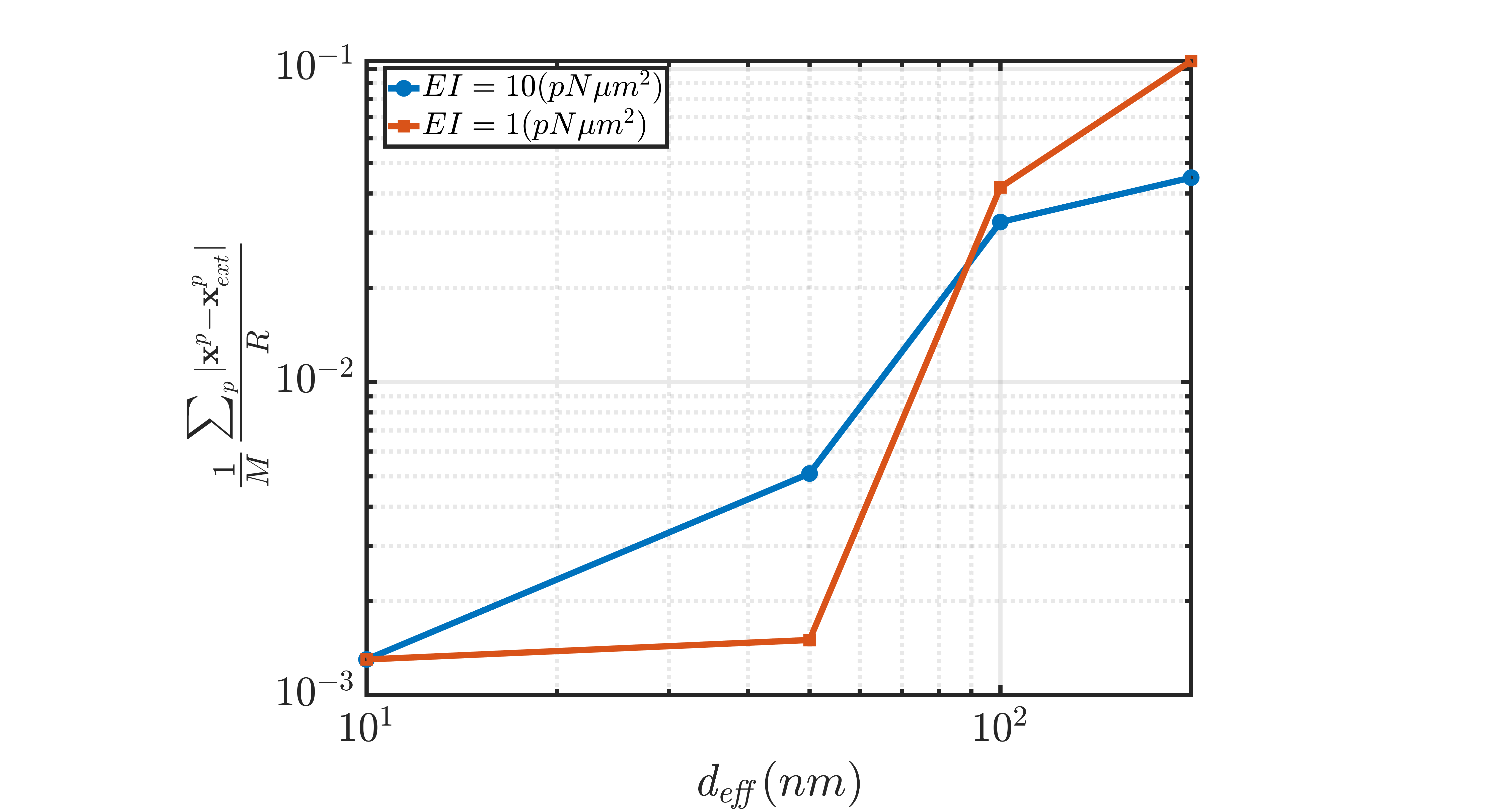}
	\caption{The maximum error during $0<\omega t<15$ between shapes of the extensible model and the inextensible model for varying effective filament diameter and two bending rigidities, $EI = 1$ and $10~ \rm pN \mu m^2$. This figure suggests that for desired errors $<1\%$ of the filament radius, we can use the inextensible model as long as ${d_{\rm eff}}<50 \rm nm$.}
	\label{fig:inxshape}
\end{figure}

\textcolor{black}{
\subsubsection{Comparison to inextensible Euler-Bernoulli elasticity}
Euler-Bernoulli elasticity has previously been used to study dynamics and interactions of straight filaments in viscous flow \cite{Tornberg2004,Chakrabarti2019,Liu2018,Manikantan2015,Chakrabarti2019a}. Euler-Bernoulli theory is a special case of the general Kirchhoff rod model and can be derived from the Kirchhoff rod theory by neglecting twist elasticity and
assuming zero intrinsic curvature.   We validate our inextensible Kirchhoff rod model by comparing its results to the shapes obtained using Euler-Bernoulli theory for the same hydrodynamic forces, in a scenario with only two-dimensional bending deformations. }

\textcolor{black}{
The Euler-Bernoulli theory describes the deformed shape of the filament $\mathbf{X}(s)$ along its arclength $s$
in the presence of external force distribution $\mathbf{f}(s)$ as
\begin{equation}
\mathbf{f}(s)=-\frac{\partial}{\partial s}(T \frac{\partial \mathbf{X}}{\partial s})+EI \frac{\partial^4 \mathbf{x}}{\partial s^4}.
\label{eq:EB}
\end{equation}
Here $T(s)$ is a tension along the filament which must be solve for to satisfy the inextensibility constraints $|\frac{\partial \mathbf{X}}{\partial s}|=1$ of the filament \cite{Tornberg2004}. To compare our Kirchhoff rod solutions to the Euler-Bernoulli elasticity, we compared deformed shapes of a cantilever beam with circular cross-sections under uniform flow $U$ past it as shown in Fig. \ref{fig:EB}a. The diameter $d$ and length $L$ of the beam are same as the filament’s diameter and arclength used before for the helical filament and the initially straight cantilever beam lying along the $x$-direction with centerline positions $\mathbf{X}_0(s)$ is discretized into $100$ segments along its arclength. Two different flow velocities $U=10,~\rm{and}~100 (\mu m/s)$ are used for these comparisons corresponding to the non-dimensional parameters $\mu U L^3/EI=1~\rm{and}~10$, respectively, which are in the range that real organisms experience swimming in viscous flow. To obtain deformed shapes from the Kirchhoff rod model $(\mathbf{X}^{KR})$, first we solve our inextensible model and calculate for the deformed shape and external force distribution $\mathbf{f}(s)$ at different non-dimensional times $Ut/L$ until the steady-state deformation is reached. Then, this force distribution is used in the Euler-Bernoulli theory Eq. (\ref{eq:EB}) to obtain the corresponding deformed shape $(\mathbf{X}^{EB})$ at these times. Boundary conditions for Eq. \ref{eq:EB} are $T|_{s=L}=0,~ \partial^2 \mathbf{X}/\partial s^2 |_{s=L} = \partial^3 \mathbf{X}/\partial s^3 |_{s=L} = \mathbf{0}$ at the free end (zero forces and torques) and $\mathbf{X}|_{s=0}=\mathbf{0},~\partial \mathbf{X}/\partial s|_{s=0}=\hat{\mathbf{x}}$ (zero displacement and rotation at $s=0$). In Fig. \ref{fig:EB}b and c, we compare the deformed shapes of the inextensible Kirchhoff rod (circles) and Euler-Bernoulli beam (solid lines) for different flow velocities $\mu U L^3/EI=1~\rm{and}~10$ at different times.  Visually, the agreement is quite good, especially for small deflections.  We quantify the error in Fig. \ref{fig:EB}c.  The curves associated with the left axis show the magnitude of the difference in displacements between our Kirchhoff rod and the Euler-Bernoulli theory, $|\mathbf{X}^{KR}(s) - \mathbf{X}^{EB}(s)|$, which are largest at the free end, where they are less than $0.024L$. The curves associated with the right axis show the relative percentage error, $E(s) = 100 \frac{|\mathbf{X}^{KR}(s) - \mathbf{X}^{EB}(s)|}{|\mathbf{X}^{KR}(s) - \mathbf{X}_{0}(s)|}$, which is less than 2.6\% at the free end.  The error is large at the fixed end because the denominator in the error becomes small, even though the difference in displacements is small.  The reason that the Kirchoff rod model differs from the Euler-Bernoulli theory for larger deflections is that in our discretization of the filament, each finite segment may have torques exerted on it by the fluid, which are not accounted for in the Euler-Bernoulli theory. 
}

\begin{figure}[h]
	\centering
	\includegraphics[width= \linewidth]{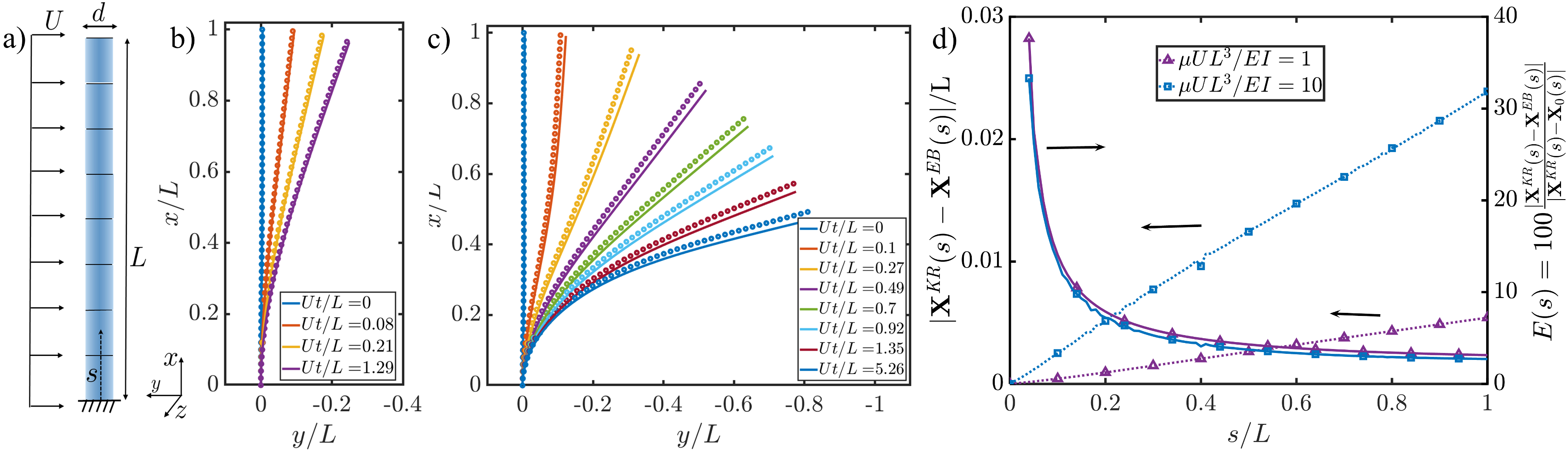}
	\caption{Comparison of the inextensible Kirchhoff rod model to the inextensible Euler-Bernoulli beam theory. a) Schematic representation for a cantilever beam of length $L$ and diameter $d$ in a uniform flow $U$. The initially straight beam is in $x$-direction and uniform flow applied in $y$-direction perpendicular to its initial direction.
	b,c) Deformed shape of the beam for different times $Ut/L$, for $\mu U L^3/EI=1~\rm{and}~10$, respectively. Circles are from the inextensible Kirchoff rod model; solid lines are from the Euler-Bernoulli theory. The most deflected final shapes in each plots are the steady-state beam deflections. 
	d) The absolute errors (dashed lines, left axis) show small differences between the two models with greatest error at the free end, and the local percentage error (solid lines, right axis) at the free end of the beam is about $2.6\%$. The percentage error is large at the base, even though the absolute error is small, since the magnitude of the deflection is small.}
	\label{fig:EB}
\end{figure}
}

\subsection{Force and Torque on the filament for different approaches}
\begin{figure}[!t]
	\centering
	\includegraphics[width=0.9\linewidth]{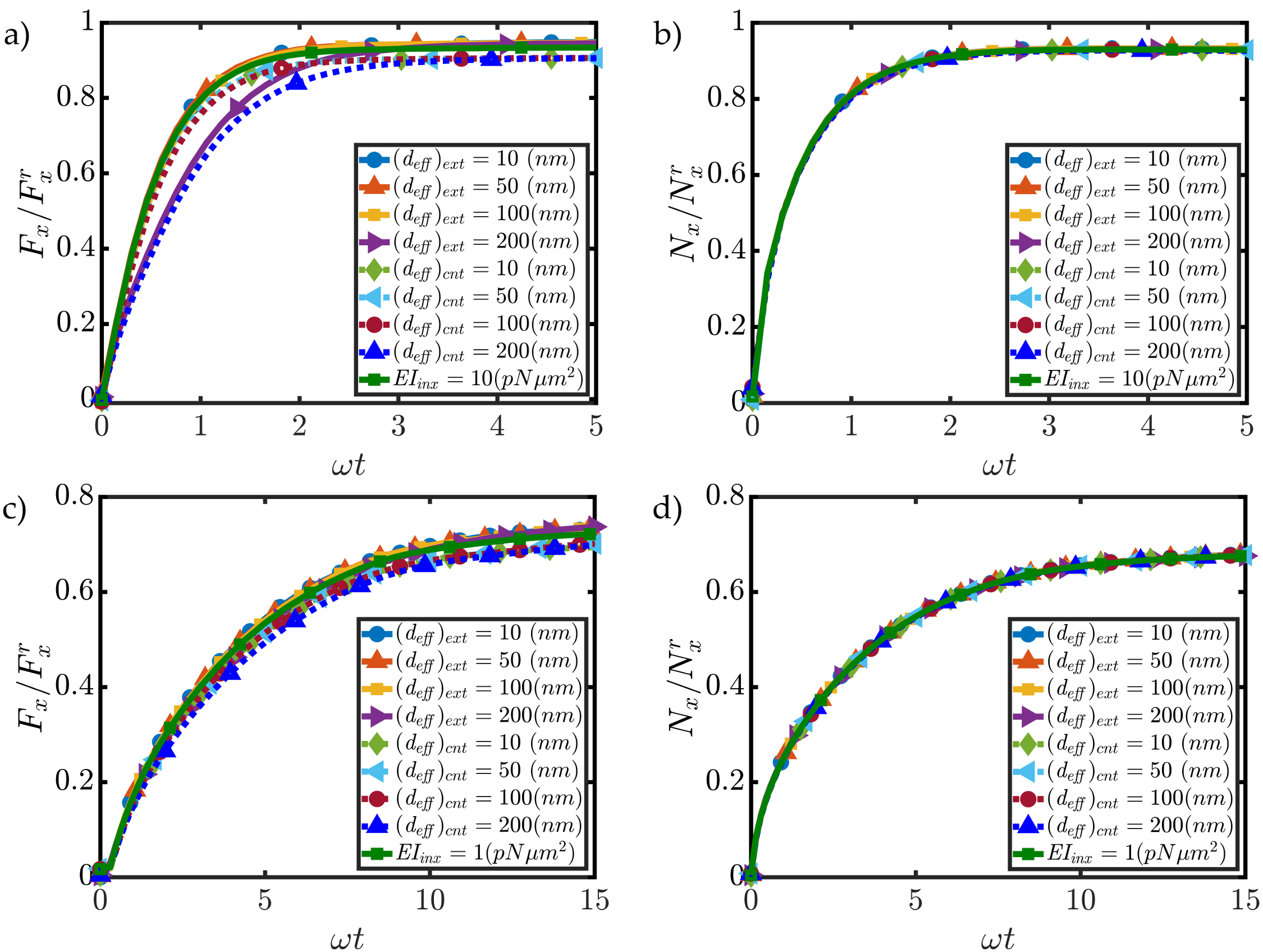}
	\caption{ 
		Comparing forces and torques predicted from different numerical approaches. The normalized force and torque component in the prescribed rotation direction ($x$-direction) are plotted. Results normalized by $N^r_x$ and $F^r_x$, the torque and force in the $x$-direction for rigid body motion of the filament in its rest configuration and the same prescribed rotation rate. Torque and force for a filament with bending rigidly a,b) $EI = 10~ \rm pN \mu m^2$ and c,d) $EI = 1~ \rm pN \mu m^2$ for different stretch ratios. }
	\label{fig:force-torque}
\end{figure}

Another measure of the accuracy of the different approaches can be obtained by comparing the total force and torques 
on the filaments.
We plot the $x-$component of the forces and torques ($F_x, N_x$) at the base  of filament ($s=0$) over time in Fig. \ref{fig:force-torque}. The filament is aligned in the $x-$direction for its rest configuration at $t=0$ as described in \S \ref{sec:geometry} and has zero velocity and a prescribed rotation rate at its base. The force and torque components are normalized by the values calculated for a rigid body with the undeformed filament shape and the same prescribed rotation rate. 
From Fig. \ref{fig:force-torque}, we see that the force and torque calculated from the extensible and inextensible models are very close to each other.

For the centerline distribution of Stokeslets, the results depend on the choice of blob parameter.  There is no systematic way in which the results converge as blob parameter is varied, since it must approximately represent the thickness of the filament. In our calculations the blob parameter was chosen to match the torque of rigid helices, so the torques in Fig. \ref{fig:force-torque}b,d are all quite similar.  However, the forces calculated using the centerline distribution in Fig. \ref{fig:force-torque}a,c have about  $5\%$ errors. We note this is significantly better than the results for centerline distributions when rotlets and torques are not included \cite{Martindale2016}, but there are still 5\%-20\% errors in deformed shapes as reported in \S\ref{sec:accuracy}.

\subsection{Computational time and efficiency}
\begin{figure}[!t]
	\centering
	\includegraphics[width=0.9\linewidth]{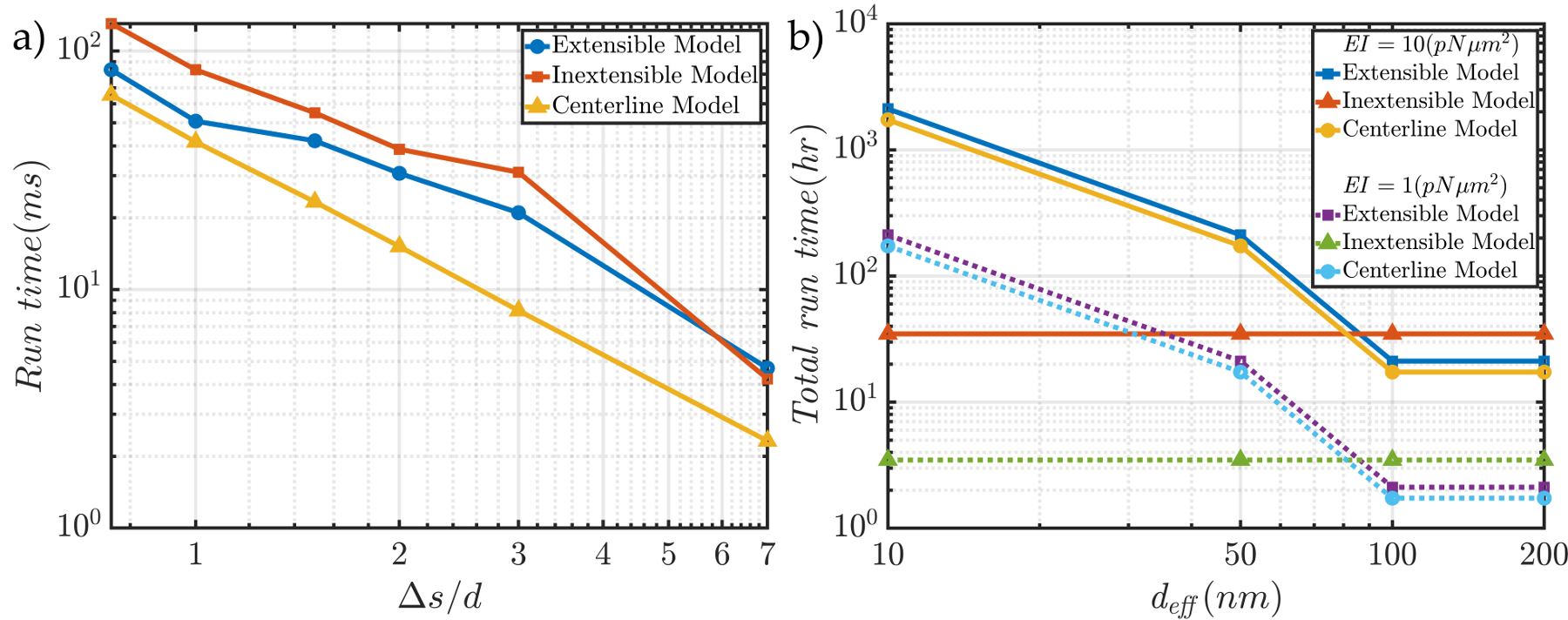}
	\caption{Log-log plot of the execution time of different numerical approaches for a) a single timestep for varying segment sizes, and b) total solution times for $0< \omega t<15$.}
	\label{fig:runtime}
\end{figure}

Finally, we compare the computational expense of the different methods discussed in this report.
We expect that the longer time-steps allowed by the inextensible approach should decrease the overall computational time needed. 
To calculate run times, numerical experiments are run in MATLAB (version 2019a) using \texttt{tic} and \texttt{toc} command pair on an Intel Core i7-6700 CPU. The average run time for a time-step, averaging over time-steps that update and do not update hydrodynamic interactions, is plotted in Fig. \ref{fig:runtime}a for different numerical approaches and segment sizes. The execution time for different approaches are in the same order of magnitude. 
Since the centerline distribution has the least number of regularized Stokeslets it is fastest, but only by a factor of about 2.  

The total computational time needed for a fixed simulation time depends on both the computational time per time step and the number of time steps needed. 
In Fig. \ref{fig:runtime}b, we plot the total time needed to simulate $t=15/\omega$ of rotation for each method, for $\Delta s/d=1$ and various effective filament diameters. 
The number of time-steps needed is determined by dividing the total simulation time by the appropriate $\delta t_0$  described in \S\ref{sec:time-conv}.
The small difference for the total run times between the centerline and extensible models is due to the increased expense of the hydrodynamic update for the larger number of regularized Stokeslets in the surface distribution.
The longer time-steps allowed by the inextensible approach for ${d_{\rm eff}}<50 \rm nm$ make it clearly more efficient in those cases. 

\section{Discussion and conclusions}
In this paper, we have studied numerical approaches to solve fluid-structure interactions of slender filaments in viscous flows. We discussed the stretching and bending timescales needed to resolve the dynamics of filaments. We showed that these timescales depend on the segment sizes, filament diameter, and Young's modules. For thin filaments ($\textcolor{black}{d_{\rm eff}}<50 \rm nm$, see Table \ref{table:EAEI}), the stretching timescale is much less than the bending timescale and requires the use of smaller time-steps, motivating our development of an inextensible approach that eliminates stretching dynamics.  We showed that for filaments with ${d_{\rm eff}}<50 \rm nm$  (or equivalently for the stretching/bending ratios of ${EA/EI}>10^4~ \rm 1/\mu \rm m^2)$, our inextensible model is both accurate and faster when compared with extensible models. Thus our inextensible model should be useful to study the dynamics of many filaments described in Table \ref{table:parameters} such as prokaryotic flagella, microtubules, and actin filaments.  
\textcolor{black}{In this manuscript, we used explicit time integration schemes to test the relative efficiency of the inextensible and extensible models.  Implicit time integration can also be used to further ameliorate issues arising from stiff numerics in both models.  However, typically implicit methods require solution of an algebraic equation involving the degrees of freedom of the problem, and because the inextensible model has fewer degrees of freedom than the extensible model, the inextensible model is also more advantageous for implicit time-stepping, in addition to the effects of eliminating the stiffest extensional degrees of freedom we have investigated here.}

\section*{Acknowledgments}
We acknowledge support from CBET-1805847 and CBET-1651031 to HCF and the University of Utah Center for  High Performance Computing.

\bibliography{refs}

\end{document}